\documentclass{article}
\usepackage{afterpage}
\usepackage{jheppub}

\usepackage{amsmath,amsfonts,amsthm,graphicx,upgreek}        

\usepackage{hyperref}

\newcommand{\gray}{\color{gray}}

\usepackage{color}

\definecolor{gray}{rgb}{0.5,0.5,0.5}
\usepackage[normalem]{ulem}

\def\th{\theta}        
\def\a{\alpha}   
\def\s{\sigma}        
        
\def\tp{{p}}   
\def\tth{{\tilde{\th}}}   
\def\L{L}

\def\<{\langle}        
\def\>{\rangle}

\def\i2{\frac{i}{2}}        
\def\cN{{\cal N}}        

\def\tP{~ 10^}
\def\TP{10^}

\newcommand{\HyrXiv}[1]{\href{http://www.arxiv.org/abs/#1}{{[{\tt arXiv}~:~{\tt
 #1}]}}}
     
\newcommand{\iq}{{\frac{i}{ 4}}}        
        
\newcommand{\id}{{\frac{i}{2}}}        
    
\newcommand{\MyPi}{{\textrm{\boldmath{\(\Pi\)}}}}

\title{ Finite Size Spectrum of         
\(SU(N)\) Principal Chiral Field  from Discrete Hirota Dynamics         
}        
     
\author[a,b]{~~~Vladimir Kazakov%
\note{member of Institut Universitaire de
    France}}
\author[a,c,d]{~~~Sebastien Leurent}

\affiliation[a]{Ecole Normale Superieure, LPT,  75231 Paris CEDEX-5,
  France}
\affiliation[b]{Universit\'e Paris-VI, Paris, France}
\affiliation[c]{Imperial College,
London SW7 2AZ, United Kingdom}
\affiliation[d]{Institut de Math\'ematiques de Bourgogne, UMR 5584 du CNRS, Universit\'e de
  Bourgogne, 9 avenue Alain Savary, 21000 DIJON, France}

\emailAdd{kazakov$\bullet$lpt.ens.fr}
\emailAdd{leurent$\bullet$nsup.org}

\abstract{Using  recently proposed method of discrete Hirota dynamics for  integrable (1+1)D quantum field theories on a finite space circle of length \(\L\)         
we derive and test numerically a finite system of nonlinear integral equations
for the exact spectrum   of energies of \(SU(N)\times SU(N)\)   
principal chiral field model as functions of \(m\L\), where \(m\) is the mass scale. We propose a determinant solution of the underlying   
Y-system, or Hirota equation, in terms of Wronskian determinants of   
\(N\times N\) matrices  parameterized by \(N-1\) functions of the   
spectral parameter \(\th\) with the known analytic properties at finite   
\(\L\). Although the method works in principle for any state, the   
explicit equations are written for states in the \(U(1)\) sector   
only. For \(N>2\), we encounter and clarify  a few subtleties in these   
equations related to the presence of bound states, absent in the   
previously considered \(N=2\) case.  As a demonstration of efficiency of our method, we solve these equations   
numerically for a few low-lying states at \(N=3\) in a wide range of   
\(m\L\).}        
        
\subheader{LPT ENS-10/25}

\newenvironment{DIFnomarkup}{}{}

\DeclareMathOperator{\im}{Im}
\usepackage{varioref}

\begin{document}

\maketitle

\section{Introduction}        

Integrable \(1+1\) dimensional  quantum field theories on a finite  
space circle   have been  rather intensively  studied in the last 20 years    \cite{DdV,Zamolodchikov:TBA1990,Bazhanov:1996aq,Dorey:1996re,Fioravanti:1996rz,Balog:2003yr,Hegedus:2004xd,Bazhanov:1994ft,Teschner:2007ng,Fioravanti:2007un}. A 
 great deal of success in the exact treatment of the finite size  
 effects in various integrable QFT's  was due to the  thermodynamic Bethe  
 ansatz (TBA) approach \cite{Tsvelik-Wiegmann}  resulting in a  system (in most of the cases  
 infinite)  of non-linear integral equations. It was realized that the  
 TBA equations could be rewritten  in a functional, Y-system form  
 \cite{Zamolodchikov:TBA1990}.       

Recently, a novel,  quite general approach to these  
problems was proposed in \cite{Gromov:2008gj} based on the  
integrability of the Y-system. The Y-system is known to be a gauge  
invariant version of the famous Hirota bilinear equation, often called  
the T-system, in its discrete form\cite{Kuniba:1993cn}. The underlying
discrete Hirota   
dynamics is integrable and general solutions of Hirota equations can be found in a  
determinant (Wronskian) form \cite{Krichever:1996qd} for various boundary  
conditions corresponding to a variety  of different  problems, from  matrix models to  
quantum spin chains and  quantum sigma-models.      For finite rank  
symmetry groups,  the Wronskians  contains only a finite  
number of functions of the spectral parameter. Thus the Wronskian solution can drastically simplify  
the problem: the infinite Y-system is reduced to a finite number of  
non-linear integral equations for these functions. Then the subtlest  
point comes: We should guess the analytic properties of these  
functions w.r.t. the spectral parameter. This is relatively easy to do  
for the spin chains where the polynomiality of transfer matrix  leads  
to the final answer in terms of a set of  Bethe ansatz equations. For  
the QFT's at a finite volume \(\L\)   (length  of the space circle) the  
situation is much more complicated and the analyticity properties of  
the Y-functions are not so obvious.  Nevertheless, it often appears to  
be possible to extract them, partially from physical considerations,  
partially from certain assumptions of the absence of unphysical  
singularities.  It helps to transform the Y-system into a system of  
non-linear integral equations (NLIE), more tractable, and better  
suitable for the numerical studies. The resulting equations can remind  
the Destri-DeVega NLIE  \cite{DdV} or even coincide with them for a limited set of 2d QFT's where these DDV equations are known.         
    
This program was first performed in \cite{Gromov:2008gj} for the       
\(SU(2)_L\times SU(2)_R\)   principal chiral field (PCF) for a general quantum state, and the       
numerical study of the finite size spectrum was successfully done  for a       
variety of interesting states, from the vacuum and mass-gap to   quite       
general states, in the so called \(U(1)\) sector or even lying out of  it (i.e. having       
excitations in left and right \(SU(2)\) spin modes).   

 In this paper, we will       
construct within these lines the  corresponding NLIE's for       
\(SU(N)\times SU(N)\)   PCF at any \(N\). We use the Wronskian       
solution of \cite{Krichever:1996qd}  for the underlying Hirota       
equation in terms of  determinants of \(N\times N\) matrices and       
guess the correct analytic form of the functions entering the       
Wronskian. For the vacuum state, the asymptotic Bethe ansatz (ABA)  
based on the scattering theory and, strictly speaking, valid only for  
sufficiently large length \(\L\) teaches us that  there are no  
singularities on the        
physical strip of the rapidity plane, at least for not too small   
\(\L\)'s.\footnote{This argument based on ABA cannot exclude a possibility that at a sufficiently small  
  size, some extra singularities occur. However, our numerics give serious evidence  
  that at least for \(N=3\) such extra singularities do not appear.}

 For excited states there are certain poles entering the physical   
 strip, and their  qualitative structure can be guessed from the ABA.  The   
 explicit construction  is done only for states in the \(U(1)\)   
 sector, but we  sketch out the generalization to any state. We show   
 how the exact S-matrix of the model (including the CDD factor)   
 naturally emerges from this approach based on the Y-system by simple analyticity   
 assumptions.   
    
The presence of additional singularities on the physical strip related to the bound states, absent for \(N=2\),        
leads in \(N>2\) case to  significant modifications, already in the expression for the energies of excited states. We find from our NLIE's the finite size (L\"uscher) corrections        
which reveal the presence of the so called \(\mu\)-terms.        
We also test our  NLIE's analytically, comparing the results with the  
known analytic data in the ultraviolet (conformal) limit. Finally,  
we demonstrate the  power of our approach by solving the resulting NLIE's numerically, for the vacuum energy and the energies of some low lying excited states  as  
functions of the size \(m\L\)  for \(N=3\).          
    
One of the principal motivations for our work was  the possibility to realize  
the same program in the case of recently constructed AdS/CFT Y-system  
\cite{Gromov:2009tv,Gromov:2009bc,Gromov:2009zb} for the exact
spectrum of anomalous dimensions in   
N=4 supersymmetric Yang-Mills theory. The PCF model, having \(N-1\)  
particles (including \(N-2 \) bound states) in its asymptotic  
spectrum, bears many similarities with the AdS/CFT case where the  
number of bound states is infinite. The corresponding Wronskian  
solutions of AdS/CFT Y-system, or Hirota equation with the so called  
T-hook boundary conditions is also available \cite{Gromov:2010vb,GKLT}.

\section{The  principal chiral field model in the large volume }   

In this section we will give the definition of the PCF model, remind  
the reader the basics of scattering theory for the physical particles and the  
ABA equations, and describe the equations for the  finite size spectrum in terms of the  
Y-system.          
    
\subsection{ The PCF model, its S-matrix and the large \texorpdfstring{$L$}{L} ABA}

The   \(SU(N)\times SU(N)\)   PCF        
model has the classical action  
\begin{align}
\mathcal{\mathcal{S_{\text{PCF}}}}=&-\frac{1{}}{2\,e_{0}^{2} } \int{\mathrm{d}\tau}
\,\int _0^\L {\mathrm{d}\sigma} \:\ {\rm
  tr}%
{\left[\vphantom{h_{}^{_{}}}\right.}(h^{-1}\partial_{\alpha} h_{}^{_{}})^{2}%
{\left.\vphantom{h_{}^{_{}}}\right]}\,,&
h%
{(\sigma,\tau)}\in& SU(N)\,.
\label{eq:ActionDef}
\end{align}

The spectrum of this asymptotically free  theory in the infinite volume \(\L\to\infty\)        
consists of  \(N-1\) physical particles with masses          
\begin{equation}        
m_a=m\, \frac{\sin\frac{\pi a}{N}}{\sin\frac{\pi }{N}}         
\end{equation}        
where the lowest mass scales with the bare charge \(e_0\) according to the asymptotic freedom \(m= \frac{\Lambda}{e_0}        
e^{-\frac{4\pi }{Ne_{0}^{2}}}\)(  \(\Lambda\) is a cut-off). Its wave        
function transforms in the fundamental representation under each of the  \(SU(N)\)        
subgroups. The exact S-matrix for bi-fundamental particles, found from       
the conditions of  factorizability, crossing, unitarity, analyticity and the bound       
state structure \cite{Zamolodchikov:1978xm},  reads\footnote{In the  
  \(N=2\) case, these definitions give \(%
{\breve\chi}_{_{CDD}}=-1\) corresponding to the  multiplication of \(S_0\) by \(i\).  
}       
\cite{Berg:1977dp,Wiegmann:1984ec}:   
\begin{equation}\label{eq:Smatr}        
\hat {S}_{12}(\theta) =%
{\breve\chi}_{_{CDD}}(\th)\cdot S_0(\theta) \frac{\hat       
  R_{L,R}(\theta)}{\theta-i} \otimes S_0(\theta)  \frac{\hat       
  R_{L,R}(\theta)}{\theta-i}   
\end{equation}        
\begin{equation}   
\label{eq:DresPhas}        
 S_0(\theta)=       
\frac{\Gamma \left( i\frac \theta N \right)       
\Gamma \left( \frac {1-i \theta} N \right)}       
{\Gamma \left(-i\frac \theta N \right)       
\Gamma \left( \frac {1+i \theta} N \right)}   \quad , \qquad   
{\breve\chi}_{_{CDD}}=\frac{\sinh (\pi \theta/N+i\pi/N)}{\sinh (\pi \theta/N-i\pi/N)}   
\end{equation}   
where we introduced  the standard \(SU(N)\) R-matrix \(        
\hat R_{L,R}(\theta)=\theta+i \hat P_{L,R}        
\) and \(\hat P\) is the permutation operator exchanging the left/right spins of the scattering         
particles.  In particular, crossing and unitarity lead to the following        
identity        
\begin{equation}        
\prod_{k=-\frac {N-1}{2}}^{\frac {N-1}{2}}S_0(\theta+ik)={-} \frac{\theta-i \frac{N-1} 2}{\theta+i \frac{N-1} 2}       
\label{crossing}        
\end{equation}        
on the scalar (dressing) factor.         
    
We can use this S-matrix to study the spectrum        
of \(\cN\) particles on a periodic space circle of a sufficiently big        
circumference \(\L\gg m^{-1}\)   imposing          
periodicity of the wave function         
\begin{equation}        
\prod^{N}_{j=k+1 }{\cal \hat S}(\theta_k-\theta_j)        
\prod^{k-1}_{j=1 }{\cal \hat S}(\theta_k-\theta_j) |\Psi\rangle=e^{-im\L\sinh(\pi\theta_k)}|\Psi\rangle\;, \label{perintro}        
\end{equation}        
which quantizes the momenta of the physical particles. The asymptotic       
spectrum is then        
given by        
 \begin{equation}\label{assE}        
 E\simeq\sum_{j=1}^{\cN} m \cosh\left(\frac{2\pi}N\theta_j\right) +O(e^{-m\L})\,   
 \end{equation}        
where \(\theta_j\) are given by solutions to the system of nested Bethe   
equations following from the diagonalization of \eqref{perintro}. This   
diagonalization can be performed by means of the algebraic Bethe   
ansatz  and leads  to the asymptotic Bethe ansatz    
(ABA) equations  \eqref{eq:BAE} and \eqref{eq:AUX} \cite{Sutherland:1975vr,Kulish:1981gi}.\footnote{In what follows we will measure all        
dimensional quantities in the units of the mass \(m\), so that we put everywhere \(m=1\). The only continuous parameter of the problem is now the volume \(\L\).}

We will rederive the ABA equations as a large \(\L\)  limit of the Y-system  of   
the model - a system of equations valid at any finite volume \(\L\) and  presented in the next subsection. The eq. \eqref{eq:BAE} represents the diagonalized version of the   
periodicity condition  \eqref{perintro}. Eq.~\eqref{eq:AUX} is the set of   
\(2(N-1) \) nested Bethe equations for the auxiliary right and left   
magnon roots \(u_j^{(k)}\) and \(v_j^{(k)}\) following from a regularity   
condition, as we will see in section \ref{sec:TSpinChain}.   
    
Note that the ABA equations \eqref{eq:AUX} remind the Bethe ansatz equations for two inhomogeneous \(SU(N)\) spin chains with the  inhomogeneity parameters \(\th_j\) given by the rapidities of physical particles. Their dynamics is defined by the periodicity equation \eqref{eq:BAE}. So the large \(\L\) limit can be also called the ``spin chain limit''.   

\begin{figure}   
\begin{center}   
\includegraphics[height=5cm]{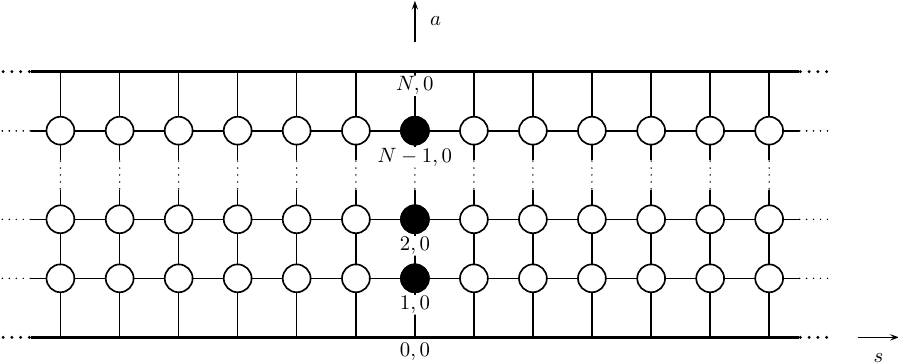}        
\end{center}  \label{fig:asStrip}        
  \caption{The \((a,s)\)-strip for Y-system and T-system}   
\end{figure}

\subsection{ TBA, Y-system and Hirota equation}        
    
The generalization of the ABA equations to any length \(\L\) is   
achieved by the TBA trick  \cite{Zamolodchikov:TBA1990}: the system   
is put on the space time torus, with a finite  space period \(\L\) and   
a big Euclidean ``time'' period \(R\to\infty\). Then, using the   
relativistic invariance, we exchange the roles of time and space and   
 solve the problem for the same system but rather with the infinite space   
extent \(R\)   with a periodic ``time" \(\L\) which can be interpreted as the   
inverse temperature \cite{Matsubara:1955ws}.   The full energy   
spectrum of such an infinite system can be found from the nested BAE   
\eqref{perintro} and from \eqref{assE} by means of the so called   
string hypothesis. The resulting equations for the densities of bound   
states are presented in  \cite{Wiegmann:PCF}, following the direct   
solution of the PCF given  in \cite{Polyakov:1983tt} (see also   
\cite{Faddeev:1985qu}). The free energy calculation at  a finite temperature for   
such an infinite volume system can be done thermodynamically, using   
the saddle point approximation due to \cite{YangYang}. Then the   
resulting integral TBA equations can be rearranged  into the Y-system\footnote{To make many formulas less bulky, the shifts   
of the spectral parameter will be often denoted as follows   
\( f^\pm=f(\theta\pm\  \frac{i}{2})\;,\,\, f^{\pm\pm}= f(\th\pm   
  i)\), and in general \(f^{[\pm k]}=f(\theta\pm\  \frac{i}{2}k)\).}   
 \begin{equation}        
\label{eq:Ysystem}        
  Y_{a,s}^+\,        
  Y_{a,s}^-=   
  \frac{1+Y_{a,s+1}}{1+(Y_{a+1,s})^{-1}}
  \frac{1+Y_{a,s-1}}{1+(Y_{a-1,s})^{-1}},\qquad \quad
  a=1,2,\dots,N-1;\quad -\infty <s<\infty
\end{equation}
where, by definition,  \(Y_{0,s}=Y_{N,s}=\infty\) and  
we have the following\footnote{Note that the notation \(\tp_a\) refers to the auxiliary model
where the roles of space and time are exchanged; in the original model,
it corresponds rather to an energy.} boundary conditions at \(\theta \to \pm\infty\):   
\begin{equation}   
\label{eq:expSupr}   
Y_{a,s}\sim e ^ {-\L \tp_a (\theta) \delta_{s,0}}\times \mathrm{const}_{a,s}\;,\qquad \tp_a= \cosh(\frac{2 \theta \pi} N )        
      \frac{\sin(\frac{a \pi}N)}{\sin(\frac{\pi}N)}\;.
\end{equation}   

This Y-system describing PCF at finite \(L\) is an infinite set of functional equations \eqref{eq:Ysystem}   
with the functions \(Y_{a,s}(\theta)\) of the spectral parameter \(\th\)  
defined in the nodes marked by black and white bullets in the interior       
of the infinite strip in  \(a,s\) lattice represented in   
fig. \ref{fig:asStrip}.

 A direct but rather tedious derivation of this Y-system was performed in the   
 Appendix A of \cite{Gromov:2008gj}  for \(N=2\). The generalization   
 of this calculation to \(N>2\) is rather straightforward, but the   
 Y-system \eqref{eq:Ysystem} is known from other considerations   
 \cite{Fateev:1991bv} and is a very universal system of equations   
 describing the integrable Hirota dynamics \cite{Krichever:1996qd}.        

As we will see later, the expression for the  momentum   
\(\tp_a(\th)\)  
is the only one compatible with the \(Y\)-system and relativistic  
invariance, up to a normalization that can be absorbed    
 into the definition of the size \(\L\) of the spin chain\footnote{So   
   that the length \(\L\) is actually measured in units of mass.}. As a result of \eqref{eq:expSupr} we see that the middle node Y-functions, \(Y_{a,0},\quad a=1,2,\cdots,N-1\), are exponentially suppressed at large \(\L\) or at large \(|\th|\).

Obviously, the Y-system  \eqref{eq:Ysystem} has  many solutions and to   
specify the physical solution we have to describe its analytic properties. To   
have a qualitative idea of the analyticity we  have to consider a   
certain limit for the solution where we know the corresponding   
Y-functions entirely as analytic functions of the spectral parameter   
\(\th\). The most convenient limit is \(\L\to\infty\) where we can solve   
the Y-system directly, with the appropriate physically natural   
analyticity assumptions,     to obtain explicitly all Y-functions and   
make a link with the exact scattering  matrix and the resulting ABA   
equations, as it was done for the \(N=2\) case in   
\cite{Gromov:2008gj}. We will give this asymptotic solution in the   
section \ref{sec:SpinChain}.  Then the Y-system, in the form of TBA equations, can be  in principle solved   
numerically by iterations, starting at large  \(\L\) and then adiabatically   
approaching \(L\sim 1\), and even very small \(\L\)'s corresponding to the   
ultraviolet CFT behavior.   The method was successfully used for   
various integrable sigma models, including the \(SU(2)\) PCF       
\cite{Gromov:2008gj,Balog:2005yz,Hegedus:2005bg,Hegedus:2004xd}. It   
will be also the main method of this paper devoted to the \(SU(N)\)   
PCF for \(N>2 \). For the vacuum state, the information from ABA is   
trivial: it  suggests that we don't have any singularities in the   
physical strip \(-iN/4<\im ( \th)<iN/4\), at least for not too small   
\(\L\)'s, since there are no Bethe roots.\footnote{It does not   
  guarantee that we will not have some singularities entering the   
  physical strip when \(\L\) becomes small enough. But our numerical result   
  don't suggest such a strange behavior.}            

The TBA procedure described above leads to the following expression for the vacuum energy         
\begin{eqnarray}        
  \label{eq:vacMenergy}        
  E_{vacuum}(\L) &=& -\frac{m}{N} \sum_{a = 1}^{N-1} \frac{\sin(\frac{a    
    \pi}{N})}{\sin(\frac{\pi}{N})} \int_{-\infty}^{\infty} \mathrm{d}\th        
\cosh\left(\frac{2 \pi}{N} \th \right) \log \left( 1 + Y_{a,0}(\th)        
\right)        \,.
\end{eqnarray}

With certain modifications in the analytic properties of Y-functions,   
described in the next section, the equations   
(\ref{eq:Ysystem}-\ref{eq:vacMenergy})  appear to be appropriate not   
only for the vacuum state, as it was originally  derived from the   
string hypothesis, but also for the excited states \cite{Bazhanov:1996aq,Dorey:1996re}. \(Y\)-functions   
for various excited states differ by their analytic properties which   
can be qualitatively inferred, as it was mentioned above, from the   
same states in the ABA. A naive heuristic proposal which worked well   
for \(N=2\) case is that the excited states correspond to the   
appearance of logarithmic poles in the integrand of   
\eqref{eq:vacMenergy} at the points \(\th_j\) where       
\begin{equation}\label{BetheL}Y_{1,0}(\th_j+ iN/4)+1=0\,.\end{equation}   
 If the contour is deformed so that it encircles these singularities,   
 the pole calculation will give a contribution \(\sum_{j=1}^\cN m   
 \cosh\frac{2\pi\th_j}{N}\)     which fits well the prediction  of the   
 ABA formula \eqref{assE}. However, the situation appears to be more   
 complicated at \(N\ge 3\), already because of the fact that unlike   
 the \(N=2\) case of \cite{Gromov:2008gj}, the solutions  \(\th_j \) of   
 \eqref{BetheL}  are not necessarily real   and this naive   
 prescription should be slightly modified in order to get a real 
 energy. This will be explained in    
 detail in the section \ref{sec:Energy}. One should admit that the right formulas for energies of excited states in the integrable sigma-models  are still rather a matter of a natural guess then of a reliable derivation. More insight is needed into this issue.

To solve the \(Y\)-system equation (\ref{eq:Ysystem}) we will often   
use it in the  form of the Hirota equation     
\begin{eqnarray}   
\label{eq:Hirota}   
  T_{a,s}^+T_{a,s}^-&=&T_{a+1,s}T_{a-1,s}+T_{a,s+1}T_{a,s-1}   
\end{eqnarray}   
 on a set \(\{T_{a,s}\}\) of functions of the spectral   
parameter \(\theta\) related to the original \(Y\)-functions as follows   
\begin{equation}   
\label{eq:DefTs}   
Y_{a,s}=\frac{T_{a,s+1}T_{a,s-1}}{T_{a+1,s}T_{a-1,s}}\,.   
\end{equation}   
On the boundary, one sets \(T_{a,s}=0\) if \(a\notin\{0,1,\dots,N\}\), so
that \(T\)-functions are associated to the nodes of the grid on figure
\ref{fig:asStrip}, including the boundaries \(a\in\{0,N\}\).

The Hirota equation (\ref{eq:Hirota}) is invariant under the gauge transformation   
\begin{eqnarray}   
\label{eq:5}  T_{a,s}&\to&   
\newcommand{\MyC}[2]{\chi_{#1}^{[#2]}}   
\MyC{1}{a+s}\MyC{2}{a-s}\MyC{3}{-a+s}\MyC{4}{-a-s}   T_{a,s}   
\end{eqnarray}   
so that \(T\)-functions are gauge dependent, whether as 
\(Y\)-functions \eqref{eq:DefTs} are gauge invariant. Another useful relation following from (\ref{eq:Hirota})   is   
\begin{equation}   
     \label{eq:1PY}   
     1+Y_{a,s}=\frac{T_{a,s}^+T_{a,s}^-}{T_{a+1,s}T_{a-1,s}}   \,.
   \end{equation}

\section{Central node equations}   

The central node Y-functions \(Y_{a,0}\) related to the black, momentum carrying  nodes on the fig.~\ref{fig:asStrip} play a special role in the Y-system. It will be useful for the future to solve the corresponding Y-system equations  for these functions entering the l.h.s. of \eqref{eq:Ysystem} in terms of the r.h.s.    

Let us rewrite  the   
\(Y\)-system \eqref{eq:Ysystem} in the form   
\begin{eqnarray}        
\label{eq:Ycentral}        
\frac{  Y_{a,s}^+        
  Y_{a,s}^-}{  \left(Y_{a+1,s}  
  \right)^{1-\delta_{a,N-1}}\left(Y_{a-1,s}\right)^{1-\delta_{a,1}}}  
&=&   
\frac{1+Y_{a,s+1}}{\left(1+Y_{a+1,s}  
  \right)^{1-\delta_{a,N-1}}} \frac{1+Y_{a,s-1}}{\left(1+Y_{a-1,s}\right)^{1-\delta_{a,1}}}  \,.
\end{eqnarray}        
 At \(s=0\) it can be rewritten  using  \eqref{eq:DefTs}-(\ref{eq:1PY})  as follows 
\begin{equation}        
\label{eq:YcentralN}        
Y^{\star \Delta}_{a,0}=\frac{T^{\star \Delta}_{a,1}(T^{(L)})^{\star \Delta}_{a,-1}}   
{T^{\star \Delta}_{a+1,0}T^{\star \Delta}_{a-1,0}} \times  
\left( \frac{T_{N,0}^+T_{N,0}^-}{T_{N,1}T_{N,-1}^{(L)}}\right)^{\delta_{a,N-1}}  
\left( \frac{T_{0,0}^+T_{0,0}^-}{T_{0,1}T_{0,-1}^{(L)}}\right)^{\delta_{a,1}}  
\end{equation}   
where we introduced a discrete D'Alembert operator \(\Delta\) on the interval \(a\in[1,N-1]\) defined   
by the formula\footnote{The terming "discrete D'Alembert operator" becomes clear if one takes the logarithm of the r.h.s. and the l.h.s. of \eqref{aaa}.}     
\begin{equation}   
F^{\star \Delta}_{a}:=   
\frac{F_{a}^+F_{a}^-}{\left(F_{a+1}\right)^{1-\delta_{a,N-1}}\left(F_{a-1}\right)^{1-\delta_{a,1}}}\label{aaa}\end{equation}   
for any function \(F_a(\th)\), and  \(\delta\) is the Kronecker  
symbol, used to add the counter terms necessary to satisfy  
\eqref{eq:Ycentral} even at \(a=1\) and \(a=N-1\).  
 By a superscript \((L)\) in   \(T^{(L)}_{a,-1}\)   
 in   \eqref{eq:YcentralN}   
we denoted  a \(T\)-function in a gauge  of a type  \((L)\)  which can be different from the gauge of the other \(T\)- functions in that formula. We can do so   
because  \(1+Y_{a,s}=\frac{\left(T_{a,s}\right)^{\star  \Delta}}{T_{0,s}^{\delta_{a,1}}T_{N,s}^{\delta_{a,N-1}}}\)  in the right hand side   
of (\ref{eq:Ycentral}) are gauge invariant, and we are allowed to   
write each \(Y\)-function in terms of \(T\)'s  taken in a different gauge. The meaning and the notation of the gauge \((L)\) will be explained later.

We can act by \(\Delta^{-1}\)on both sides of  
\eqref{eq:YcentralN}, to get  
\begin{equation}  \label{eq:YcentralInterm}
  Y_{a,0}=e^{-\L\tp_a(\th)}\frac{T_{a,1} T^{(L)}_{a,-1}}   
{T_{a+1,0} T_{a-1,0}}   
\left(   
\left( \frac{T_{N,0}^+T_{N,0}^-}{T_{N,1}T_{N,-1}^{(L)}}\right)^{\delta_{a,N-1}}  
\left( \frac{T_{0,0}^+T_{0,0}^-}{T_{0,1}T_{0,-1}^{(L)}}\right)^{\delta_{a,1}}  
\right)^{\star  \Delta^{-1}}   
\end{equation}  
where  \(\tp_a= \cosh(\frac{2 \theta \pi} N )        
      \sin(\frac{a \pi}N)/\sin(\frac{\pi}N)\).
The
factor \(e^{-\L\tp_a(\th)}\) is a zero mode of \(\Delta\), in the sense that
\(\left(e^{-\L\tp_a(\th)}\right)^{\star \Delta}=1\), and it is added in order
to reproduce the asymptotics \eqref{eq:expSupr}. This equation
\eqref{eq:YcentralInterm} is valid up to a zero mode, which will be
discussed in the next sections, though we can already see that this
remaining zero-mode has a constant  asymptotics at large \(\theta\).
Furthermore, the action of the operator \(\Delta^{-1}\) can be  
easily calculated by the discrete  Fourier transform in \(\theta,a\) variables,
so that the final expression is  
\begin{gather}   
\label{eq:YcentralR}        
Y_{a,0}=e^{-\L\tp_a(\th)}\frac{T_{a,1} T^{(L)}_{a,-1}}   
{T_{a+1,0} T_{a-1,0}}   
\left(   
\MyPi_{N-a}\left[ \frac{T_{0,0}^+T_{0,0}^-}{T_{0,1}  T^{(L)}_{%
      0,-1}} \right]   
\MyPi_{a}\left[ \frac{T_{N,0}^+T_{N,0}^-}{T_{N,1}  T^{(L)}_{%
      N,-1}} \right]   
\right)^{\star  K_N}\,,   \\
{\textrm{where } f^{\star  K_N} = e^{\log f \star  K_N}}
\end{gather}        
and \(\star \) stands for   convolution; the   
     ``fusion'' operator \(\MyPi_s\) is defined as the following product    
\begin{equation}   
  \MyPi_k[f](\theta)=\prod_{j=-(k-1)/2}^{(k-1)/2}f(\theta+i~j)
=f^{[-k+1]}f^{[-k+3]}\dots f^{[k-3]}f^{[k-1]}
\end{equation}   
and the kernel   
\(K_N\) is the operator inverse to \(\MyPi_N\): \(\forall f\,\, \mathrm{regular},   
\left(\MyPi_{N}[f]\right)^{\star  K_N}=f\). Its Fourier transform is   
\begin{equation}        
  \label{eq:KerDef}        
  \widetilde{K_N}(\omega)=\frac 1 {\sum_{j=-\frac {N-1} 2}^{\frac        
      {N-1} 2} e^{2i \pi j \omega}}\,.\end{equation}   
Back in the \(\th\)-space it takes the form 
      \begin{equation}        
K_N(\th)=\frac{1}{2 N}\left[\mathrm{tan}         
    \left(\frac{\pi-2\pi i        
        \th}{2N}\right)+\mathrm{tan}\left(\frac{\pi+2\pi i        
        \th}{2N}\right)\right]=\frac{1}{N}\frac{\sin(\pi/N)}{\cosh(2\pi \theta/N)+\cos(\pi/N)}        \,.
\end{equation}        

\section{The large \texorpdfstring{$\L$}{L}, ``spin chain''     
limit of Y-system and its relation to ABA }        
\label{sec:SpinChain}        

We will derive in this section the large \(\L\), ABA  equations   
\eqref{eq:BAE},\eqref{eq:AUX} directly from the Y-system   
\eqref{eq:Ysystem}. Following the logic of \cite{Gromov:2008gj} we use   
the fact that the Y-functions of the momentum carrying (black) nodes are exponentially small in this limit:   
\begin{equation}   
\label{eq:YcentralAss}        
Y_{a,0}=\frac{T_{a,1} T_{a,-1}}   
{T_{a+1,0} T_{a-1,0}}   
 \sim e^{-\L{\tp}_a(\th)} \,.
\end{equation}       

This implies that the two wings, left (for \(s<0)\) and right (for \(s>0)\), of the Y-system \eqref{eq:Ysystem} are almost decoupled and can be treated separately.

\subsection{Expressions for \texorpdfstring{$T$}{T}-functions in the large volume,  spin chain limit}   
\label{sec:TSpinChain}   

Eq.\eqref{eq:YcentralAss} suggests that either \(T_{a,1}\sim e^{-\L{\tp}_a(\th)}\) or      
  \(T_{a,-1}\sim e^{-\L {\tp}_a(\th)}\). Which one does so (whereas another   
  one is finite) is a matter of choice of a gauge for T-functions.

We will work with two different gauges  \((R)\) and \((\L)\), such that   
in the large \(\L\) limit we have   
\begin{equation}   
\label{eq:SpinChainDef}   
    \underbrace{T_{a,-1}^{(R)}\ll 1\qquad\;,\; T_{a,1}^{(L)}\ll
      1}_{1\leq a\leq N-1}\qquad\;,\; {T_{a,s\geq
        0}^{(R)}}\sim1\qquad\;,\;{T_{a,s\leq 
        0}^{(L)}}~\sim1   
\end{equation}   
\((R)\) will be called the ``right-wing-gauge'' and \((L)\) the   
``left-wing-gauge'', and when this  superscript will  be omitted    
it will be implicitly assumed that we are working in  the \((R)\) gauge.   

In the large \(\L\) limit, the T-functions of the left \((L)\) and right \((R)\) gauge both   
describe the same \(Y\)   
functions but (up to exponential corrections) they satisfy   Hirota equation   
restricted to the wings \(s\geq0\) (resp \(s\leq0\)). Moreover, these \(T\)-functions are   
in this limit  analytic on the whole complex plane, and   
therefore polynomial.   

 Such a solution of Hirota   
equation is well known in  applications to the fusion procedure in similar spin chain   
systems, bosonic \cite{Krichever:1996qd} or even supersymmetric \cite{Tsuboi:1997iq,Kazakov:2007fy}. First we   
parameterize \(T_{1,s}\) in terms of \(N\) functions \(X^{(W)}_{(j)}(\th),\quad\ j=1,\dots,N\) by   
means of the  following  generating functional     

\begin{eqnarray}   
\label{eq:defW}   
\hat W^{(W)}&=&\left(1-X^{(W)}_{(N)}(\th)\,e^{i\partial_\th}\right)^{-1}   
\left(1-X^{(W)}_{(N-1)}(\th)\,e^{i\partial_\th}\right)^{-1}   
\dots\left(1-X^{(W)}_{(1)}(\th)\,e^{i\partial_\th}\right)^{-1}\\   
&=&\sum_{s=0}^\infty\,\frac{T_{1,%
{\pm}s}^{(W)}(\th+\id  
  (s-1))}{\varphi(\theta-N\iq)} e^{is\partial_\th}   
\end{eqnarray}         
where the   
superscript \(W=R,L\) indicates the wing that we study (either right or left), and \(\pm s\) is equal to \(s\) for the right wing (if \(W=R\)) and to \(-s\) for the left wing (if \(W=L\)). 

These functions \(X^{(W)}_{(j)}(\th)\) can be further expressed as follows   
\begin{equation}   
X^{(W)}_{(k)}=\frac{{Q^{(W)}_{k-1}}^{[N/2-k-1  
    ]}}{{Q^{(W)}_{k-1}}^{[N/2-k+1]}} \frac{{Q^{(W)}_{k}}^{
[N/2-k+2]  
  }}{{Q^{(W)}_{k}}^{[N/2-k]}}  
\,,\quad k=1,2,\dots,N     
\label{eq:DefMon}\end{equation}      
in terms of some Q-functions\footnote{The present Q-functions \(Q_{k}\)
  correspond to the functions  \(Q_{1,2,\dots,k}\) in the Hasse diagram
  notation of \cite{Tsuboi:2009ud}.} denoted as \(Q^{(W)}\). 
These \(Q\) functions
 in the corresponding gauge are polynomials  characterizing   different 
solutions of  Hirota equation in the large \(\L\) limit, their roots
are the Bethe roots describing  various excited states - solutions of
the \(Y\)-system\footnote{We assume here that there exists a
    gauge such that the large \(\L\)
    limit is described by polynomial functions \(Q_k^{(W)}\).  Although it needs a better understanding from the point of view of Y-system, it is the case if we start treating the large \(L\) limit from the S-matrix by the ABA approach.}:   
\begin{gather}\label{Q-functions}   
 Q_k^{(R)}(\th)=\prod_{j=1}^{J^{(R)}_k}\left(\th-u^{(k)}_j\right),   
 \qquad Q_k^{(L)}(\th)=\prod_{j=1}^{J^{(L)}_k}\left(\th-v^{(k)}_j\right),\qquad (k=1,\cdots,N-1)\\ 
Q_N^{(R,L)}(\th)\equiv\varphi(\th)=        
\prod^{\cN}_{j=1  }(\th-\th_j),\label{phi-function}   
\qquad   
Q_0^{(R,L)}(\th)\equiv 1   \,.
\end{gather}

In particular, we have from \eqref{eq:defW}   
\begin{eqnarray}   
\label{eq:T11}   
T_{1%
{,\pm}1}^{(W)}(\th)   
=\varphi(\theta-\frac{iN}{4})\sum_{k=1}^N\,X^{(W)}_{(k)}(\th)\,.
\end{eqnarray}     

\(T_{1,%
{\pm}1}\)  should be free of poles, i.e. polynomial. But for each  
Bethe root \(w_j=u^{(k)}_j\) or  \(w_j=v^{(k)}_j\), the two functions  
\(X_{(k)}^{(W)}\) and \(X_{(k-1)}^{(W)}\) have a pole at the same position \(w_j-\frac i  
2\left(\frac N 2 -k\right)\). By requiring their cancellation in the sum  
\eqref{eq:T11}, we get a constraint on the position of  
\(w_j\), which we will call the auxiliary  Bethe equation:  

\begin{equation}   
-1=\frac{Q_{k-1}^{(R/L)}(w_j-i/2)}{Q_{k-1}^{(R/L)}(w_j+i/2)} \frac{Q_{k}^{(R/L)}(w_j+i)Q_{k+1}^{(R/L)}(w_j-i/2)}{Q_{k}^{(R/L)}(w_j-i)Q_{k+1}^{(R/L)}(w_j+i/2)}   
\,,\quad  
\left\{\begin{array}{l}  
k=1,2,\dots,N-1\\  
w_j=u_j^{(k)} \textrm{~resp~} v_j^{(k)}  
\end{array}\right.  
\label{eq:AUX}\end{equation}

The rest of the T-functions in the right wing can be expressed through the   
Cherednik-Bazhanov-Reshetikhin (CBR) determinant\footnote{Equation
  \eqref{eq:CBR} makes sense if \(a\ge 1\), and can be extended to \(a=0\)
  under natural conventions: one can use the convention that
\(\MyPi_{-a}[f]=1/\MyPi_{a}[f]\) -- consistently with
\(\MyPi_{a+1}[f]=f^{[a]} \MyPi_{a}[f^-]\) -- and that
the determinant of the empty matrix is equal to one, so that at \(a=0\) the
relation \eqref{eq:CBR} reduces to \(T_{0,s}=\varphi^{[\mp
  s-N/2]}\).

Also note that the sign of the shift in the denominator \(\varphi^{[\mp
  s-N/2]}\) is different for \(T^{(R)}\) and \(T^{(L)}\). This sign is
actually a convention which can be fixed using the gauge freedom \eqref{eq:5}.
}
 \cite{Cherednik,Bazhanov:1989yk}     
\begin{equation}   
T_{a,s}=   
\frac{\det_{1\leq j,k\leq a}\,T_{1,s+k-j}\left(\th+\id(a+1-k-j)\right)} 
{\MyPi_{a-1}\left[
\varphi^{[\mp s-N/2]}   
  \right]}   
\label{eq:CBR} \end{equation}   
 and  they are also automatically polynomial in virtue of   
\eqref{eq:AUX}.

Among these \(Q\)-functions, the polynomial function \(Q_N=\varphi\),
encoding, as its roots,  the rapidities    
of all physical particles, will be of a particular importance,   
and the vanishing of \(T^{(R)}_{a,-1}\) or \(T^{(L)}_{a,1}\)   due to
\eqref{eq:YcentralAss} implies   the following
asymptotics\footnote{The statement
  in (\ref{eq:phiTboundRi}) is a bit too strong and we will   
  see further that it actually only holds   
  inside some strips on the complex plane. }   
\begin{eqnarray}   
\label{eq:phiTboundRi}   
\varphi (\th)&=& \lim_{\L\to\infty}   
T_{a,0}^{(R,L)}(\th+i \frac {N-2a}4)\\   
&=& \lim_{\L\to\infty}   
T_{0,s>0}^{(R)}(\th+i \frac {N+2s}4)\;\;=\;\; \lim_{\L\to\infty}   
T_{0,s<0}^{(L)}(\th+i \frac {N-2s}4)\\   
&=& \lim_{\L\to\infty}   
T_{N,s>0}^{(R)}(\th-i \frac {N+2s}4)\;\;=\;\; \lim_{\L\to\infty}   
T_{N,s<0}^{(L)}(\th-i \frac {N-2s}4)   
\end{eqnarray}   
These relations   
translate all the zeroes \(\theta_j\) of \(Y_{1,0}(\th+i\frac {N}4)+1\), giving the roots of   
 Bethe equation \eqref{BetheL}, into the zeroes of \(T\)-functions.

\subsection{Asymptotic Bethe ansatz (ABA)}   

Now we will reproduce from the Y-system in large volume limit   
\(\L\to\infty\)    the ABA equations \eqref{eq:BAE},\eqref{eq:AUX} for   
the spectrum of energies.   In this spin chain limit,   
(\ref{eq:YcentralR}) can be employed to     
compute \(Y_{a,0}\) to the leading order, by using the   
asymptotic behaviors  (\ref{eq:SpinChainDef}).   

At this point, it is interesting to notice that the crossing relation \eqref{crossing}  
implies that up to a zero mode of %
{\(\MyPi_N\)  (i.e. up to a
  function \(Z\) such that \(\MyPi_N[Z]=1\))}\footnote{One can note
  that the sign \((-1)^{1/N}=e^{i (2\,k+1)\pi/N}\) is defined up to a
  factor \(e^{i 2\,k\pi/N}\), which is a zero mode of \(\MyPi_N\) and can
  be ignored.}
\begin{gather}
\label{eq:ExpdSC0}   
  \left(   
\frac{\varphi^+}{\varphi^-}   
  \right)^{\star  K_N}   
\;=\;\frac{\varphi^{[+2-N]}}{\varphi^{[-N]}}   
  \left(   
\frac{\varphi^{[-2N+1]}}{\varphi^-}   
  \right)^{\star    
    K_N}\;=\;\frac{\varphi^{[+2-N]}S^{[-N]}}{\varphi^{[-N]}}   
\;=\;\frac{\varphi^{[+N]}S^{[+N]}}{\varphi^{[-2+N]}}   \\
\textrm{where }\;\;\;\;\;\;\; S(\th):=\prod_j 
S_0(\th-\theta_j)\;\;\;\;\;\;\;\textrm{i.e. }S=(-1)^{\mathcal{N}/N} \left(\frac{\varphi^{[-N+1]}}{\varphi^{[+N-1]}}\right)^{\star K_N}\,.
\label{eq:2}
\end{gather}
By denoting\footnote{One can note that
  \(\varepsilon^2=(-1)^{2\mathcal{N}/N}\) is a zero mode of \(\MyPi_N\),
  which is ignored as long as we work up to a zero mode.} \(\varepsilon=(-1)^{\mathcal{N}/N}\), this gives for instance   
\begin{align}   
\label{eq:SimpPhiPi0}   
\lim_{\L\to \infty}   
\MyPi_{N-a}\left[ \frac{T_{0,0}^+T_{0,0}^-}{T_{0,1} \cdot T^{(L)}_{%
      0,-1}} \right]^{\star    
      K_N}\!\!
 =&   
 \MyPi_{N-a}\left[   
       \frac{\varphi^{[-N/2+1]}}{\varphi^{[-N/2-1]}}\right]^{\star    
       K_N}\\   
 =&   
     \MyPi_{N-a}\left[\varepsilon\frac{\varphi^{[+N/2]}S^{[+N/2]}}{\varphi^{[-2+N/2]}}\right]   
 =   
 \frac{\varphi^{[3N/2-a-1]}}{\varphi^{[-N/2+a-1]}}\,\,
 \MyPi_{N-a}\left[\varepsilon S^{[+N/2]}\right]\\   
 =&   
 \frac{\varphi^{[3N/2-a-1]}}{\varphi^{[-N/2+a-1]}}   
 \frac{\varphi^{[-N/2-a+1]}}{\varphi^{[3N/2-a-1]}}   
 \frac 1 {\MyPi_{a}\left[\varepsilon S^{[-N/2]}\right]}\,.
\label{eq:SimpPhiPiL}   
\end{align}

As a consequence, the large \(\L\) limit of equation (\ref{eq:YcentralR}) is   
\begin{gather}   
\label{eq:largeLY1}   
{  Y_{a,0} (\th)  
\sim   
e^{-\L \tp_a}\frac{T_{a,1} T^{(L)}_{a,-1}}   
{T_{a+1,0} T_{a-1,0}}}   
\frac{\varphi^{[-N/2-a+1]}}{\varphi^{[-N/2+a-1]}}   
\frac{\varphi^{[-N/2-a+1]}}{\varphi^{[-N/2+a+1]}}   
\frac 1
{\MyPi_{a}\left[\left(S^{[-N/2]}\right)^2\chi_{_{CDD}}^{[-N/2]}\right]}\,,
\\
{\textrm{where } \chi_{_{CDD}}
(\th):=\prod_j\breve\chi_{_{CDD}}(\th-\theta_j) }\,.\label{eq:3}
\end{gather}
Here, the factor  \(\chi_{_{CDD}}\) (\ref{eq:DresPhas})  was added as another zero mode, necessary   
to transform the double poles and double zeroes of \(S^2\) into the simple   
ones  \cite{Wiegmann:1984ec}.  We will also see in section   
\ref{sec:chiCDD} that  this factor arises in our Y-system formalism in 
a natural way. 

 In particular, at \(a=1\), we get the ABA equation  (periodicity condition
for the wave function):   
\begin{equation}   
\label{eq:BAE}   
  -1=e^{- i \L \mathrm{sinh}\frac{2 \pi} N\theta_j}\frac   
  {1}{\chi_{_{CDD}}(\theta_j)S(\theta_j)^2}   
\frac {Q_{N-1}^{(L)}(\theta_j-i/2)}   
{Q_{N-1}^{(L)}(\theta_j+i/2)}  \frac {Q_{N-1}^{(R)}(\theta_j-i/2)}   
{Q_{N-1}^{(R)}(\theta_j+i/2)}   
\end{equation}       
which expresses the fact that \(Y_{1,0}(\theta_j+i N/4)+1=0\)  
(here \(T_{1,1}\) was replaced by the single surviving, last term  
of \eqref{eq:T11}).   

 In conclusion, we have  shown here that the \(Y\) system implies the  familiar   
 \(ABA\)   
 equations \cite{Wiegmann:PCF,Wiegmann:1984ec}. In the next sections we will   
see how these ABA equations for the spectrum of PCF can  be generalized to any finite size \(\L\).

\section{Expressions for the energy of excited states}\label{sec:Energy}        

No complete and full proof procedure is known to generalize the formula   
\eqref{eq:vacMenergy} to the  excited states\footnote{except  for  \(N=2\) case where we know from \cite{Gromov:2008gj} the complete description of all excited states}. The analytic   
continuation  of   \cite{Dorey:1996re} with respect to the mass is difficult, if   
possible at all for a general state in our model. The   
procedure of  \cite{Dorey:1996re,Bazhanov:1994ft} claims that for the excited   
states    a set of  logarithmic poles (different for each state)   
appears under the integral  in \eqref{eq:vacMenergy}. From  
\eqref{eq:phiTboundRi} and  the ABA    
we know that at very large \(\L\),  \(T_{a,0}(\theta)\simeq\varphi(\theta - iN/4+ia/2)\), where   
\(\varphi(\theta)=\prod_j(\th-\th_j)\) is a polynomial encoding all {\it   real} roots\footnote{We consider here for simplicity only the 
  situation when the Bethe roots \(\th_j\) are real in the asymptotic 
  limit. The case when they occur in complex conjugated pairs 
  should not be very different but at the moment we did not 
  try to do it. } 
For finite \(\L\) the roots \(\th_j\) will be shifted and in general become complex. These exact Bethe roots, as opposed to the approximate ones given by   
\eqref{eq:BAE},  are  defined by the exact Bethe equations  
\(T_{a,0}(\th_j^{(a)}%
{+}i N/4%
{-}i a/2)=0\). There is a whole family of such  
roots when \(a\in [0,N]\), because  even though the two functions
  \(T_{a,0}(\th)\) and \(T_{a+1,0}(\th%
{-}i/2)\) have the same limit at
  large \(\L\), they do not necessarily have the same roots at finite size.
 Each of these roots also gives rise to two zeroes  
and two poles in the \(Y\)-functions, namely, as we see from \eqref{eq:1PY},   
\(1+Y_{a,0}(\th_j^{(a)}+i N/4-i a/2\pm i/2)=0\) and   
\(1+Y_{{a\pm1},0}(\th_j^{(a)}+i N/4-i a/2)=\infty\). Among these families  
of finite size Bethe roots, we will actually restrict  
ourselves
to the roots  
\(\th_j^{( \frac N 2)}\) for even \(N\), and \(\th_j^{( \frac{ N\pm 1}{
    2})}\) for odd \(N\).  We will argue that only those ones will   
contribute as poles caught by an integration contour.  

Separating  the logarithmic poles (where \(1+Y_a\) cancels) in the contour integral \eqref{eq:vacMenergy} should give a familiar contribution \(\sum_j\cosh\frac{2\pi}{N}\th_j\) to the energy of a finite \(\L\) state.  This appears to be the right, though not completely well understood and justified, answer for some models, including the PCF at \(N=2\) \cite{Gromov:2008gj}.   

 For PCF at \(N>2 \) this  procedure encounters another difficulty: the zeros under the logarithm in \eqref{eq:vacMenergy} appear to correspond in general to  complex Bethe roots \(\th_j\).  We have to decide what is the right   integration contour in \eqref{eq:vacMenergy} when this formula is applied to an excited state. We are not aware of any  well justified procedure for fixing the contour but we shall try to guess it on the basis of our numerical observations  and the symmetry considerations.      

In the rest of this section, we consider the formula for
excited states of the \(U(1)  \) sector - the one which corresponds to
the wave function \(|\Psi\rangle\) having the maximal value of total
spin \(S_L=S_R=\cN/2\) w.r.t.  the  \(SU(N)_R \) and \(SU(N)_L\)
symmetries. In this case  \(J^{(L,R)}_k=0 \)  and there are no
auxiliary roots in  the ABA \(Q\)-functions \eqref{Q-functions} (all
of them are equal to 1 except \(Q_N^{(R,L)}=\varphi\), see
(\ref{Q-functions}-\ref{phi-function}). In what follows,    
we shall distinguish even and odd \(N\)'s.       

\subsection{Energy of state in the  \texorpdfstring{$U(1)$}{U(1)}  sector at odd \texorpdfstring{$N$}{N}'s}   
  \begin{DIFnomarkup}
\begin{figure}        
    \centering
    \includegraphics{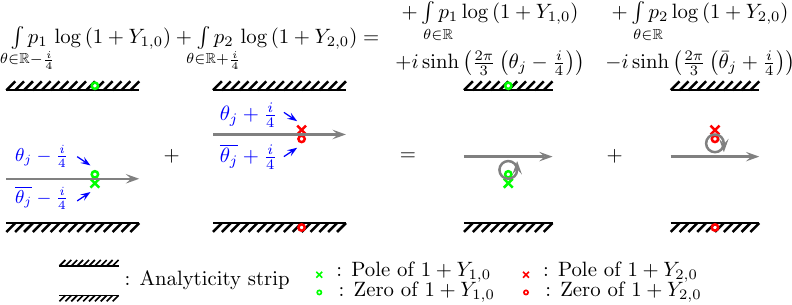}

  \caption{Analyticity of the integrand \(\cosh(\frac {2\pi}3 \th)\log\left(  
      (1+Y_{1,0}) (1+Y_{2,0}) \right)\) and manipulations with the 
    contours when \(N=3\) 
}  
  \label{fig:U1_contour}   
\end{figure}        
\end{DIFnomarkup}

It is  believed that the energy of an excited state can be obtained from (\ref{eq:vacMenergy})   
by an analytic continuation  in the parameter   
\(\L\). This continuation has the effect of appearance of new singularities of the integrand in the physical strip in (\ref{eq:vacMenergy})   
 and a certain choice of the integration contour,    
enclosing some  singularities of the integrand \cite{Bazhanov:1996aq,Dorey:1996re}. How it happens in each particular model or state is usually a rather complicated question. It implies the analysis of positions of these singularities at a finite \(\L\) but the large \(\L\)  asymptotics often serves as an important guiding principle.     

Here we  propose a  formula for the energies of excited states in the  
\(U(1)\) sector which seems to work well for any odd \(N\). It is  
based on our numerical and analytic  observations, in particular for  
the \(N=3 \) case. It reads as follows 
\begin{eqnarray}\label{eq:EnergyOdd}   
E(\L) &=& -\frac{m}{N} \sum_{a = 1}^{N-1} \frac{\sin(\frac{a        
    \pi}{N})}{\sin(\frac{\pi}{N})} \int_{-\infty-N\iq+a\id}^{\infty-N\iq+a\id} \mathrm{d}\th\,\,        
\cosh\left(\frac{2 \pi}{N} \th\right) \log \left( 1 +        
  Y_{a,0}(\th)         
\right)        
\end{eqnarray}       
so that we have the straight integration contours parallel to the real axis and shifted by \(-N\iq+a\id\).\footnote{ One of the advantages of this straight contour is that it can be easily        
 implemented in numerics. We will see indeed that the \(Y\) functions  
 can be most easily computed on exactly these lines. We will also see that the 
statement holds only for roots with even momentum number, but for odd 
momentum number, the (slightly modified) contour stays 
very close to this straight line.}    

Let us explain the  reason for such a choice of contours. First, let  
us note that in order to have a real energy from \eqref{eq:EnergyOdd}  
we should impose the following property of  Y-functions under  
complex-conjugation: \(\overline{ Y_{a,s}(\th)}=Y_{{N-a},s}( \bar \th).\)  
We will restrict ourselves to the gauges where this property is a  
consequence of the relation  
\begin{equation}  \label{eq:7}
  \overline{ T_{a,s}(\th)}=T_{{N-a},s}( \bar \th)  
\end{equation}  

For finite \(\L\), we will focus on the roots \(\th_j\) 
defined\footnote{In this section, we will denote \(\th_j\) for 
  \(\th_j^{(\frac {N-1} 2)}\), because the other types of finite size 
  roots don't contribute.} by  \(T_{\frac{N-1}2,0}(\theta_j+i/4)=0 \).  

  Due to the very definition of  
  \(1+Y_{a,0}=\frac{T_{a,0}^+T_{a,0}^-}{T_{a+1,0}T_{a-1,0}}\), each  
  \(\theta_j\) gives rise to two zeros and poles. In particular,   
\(1+Y_{\frac {N-1} 2,0}(\th)\) has a zero and a pole\footnote{In  
  addition to this zero and pole, \(1+Y_{\frac {N-1} 2,0}(\th)\) has  
  another zero at \(\theta_j+3 i/4\) and a pole at each root of  
  \(T_{\frac{N-3} 2,0}\), but this  
  will not have any consequence in the 
  contour argument.} at  
respective positions \(\theta_j-i/4\)    
and \(\overline{\theta_j}-i/4\) because   
\(T_{\frac {N-1} 2,0}^+(\theta_j-i/4)=0\) and   
\(T_{\frac {N%
{+}1} 2,0}(\overline{\theta_j}-i/4)=0\). In the large \(\L\) limit these   
zero and  pole almost coincide since \(\theta_j\) is almost real.   
By complex conjugation, we can also say that \(1+Y_{\frac {N+1}  
  2,0}(\th)\) has a zero and a pole at respective positions  
\(\overline{\theta}_j+i/4\)   and \({\theta_j}+i/4\).  

This structure is   
illustrated for \(N=3\) in figure \ref{fig:U1_contour}. From the  
L\"uscher corrections\footnote{In section \ref{sec:Luscher}, we detail  
  how this is proved in the asymptotic limit. Our numerics suggests that  
  it is still true at finite size,  
  and even in the conformal limit.}, we  
can say that the pole occurs below the zero   
for  \(1+Y_{1,0}\) and vice versa for  \(1+Y_{2,0}\), at least for  
roots with  even momentum numbers\footnote{So that the contour will  
  actually have to be slightly modified for roots having odd momentum  
  number. This will be done in such a manner that   
\eqref{eq:oddNU1energy} will stay true.  
 }. This is important to ensure the right answer if we want the contours to be straight.

 In \eqref{eq:EnergyOdd} we chose the integration contour to pass, for  
 the \(\frac {N-1} 2\) and \(\frac {N+1} 2\)'th term in the sum,
 between those zero and  pole.  Deforming the   contour to the real
 axis and computing  the contributions of the logarithmic poles
 enclosed by the contour during that deformation\footnote{The contour
   deformation is best understood after an integration by parts which
   removes logarithmic cuts and changes the \(\cosh\) into a \(\sinh\).},
 one gets the following formula        
\begin{multline} \label{eq:oddNU1energy}        
E(\L) = -\frac{m}{N } \sum_{a = 1}^{N-1} \frac{\sin(\frac{a        
    \pi}{N})}{\sin(\frac{\pi}{N})} \int_{-\infty}^{\infty} \mathrm{d}\th\,        
\cosh\left(\frac{2 \pi}{N} \th \right) \log \left( 1 + Y_{a,0}(\th)        
\right)\\        
+i \sum_j m \frac {
\mathrm{cos}\frac \pi{2 N}}       
{\mathrm{sin}\frac \pi N}\left[\sinh \left(\frac{2 \pi }{N} \left(\theta_j        
      -i/4\right)\right)-\sinh \left(\frac{2\pi}{N}         
    \left(\bar\theta_j        
      +i/4\right)\right)\right]%
\end{multline}

In the thermodynamic limit (\( \L\gg1\)), the       
Bethe roots \(\theta_j\) become real and the second line of       
(\ref{eq:oddNU1energy}) reduces to the asymptotic result  \eqref{assE}, whereas the term in the first line appears to be \(O(e^{-m\L})\).

\subsection{Energy of state in the  \texorpdfstring{$U(1)$}{U(1)}  sector at even \texorpdfstring{$N$}{N}'s}   

\begin{DIFnomarkup}
  \begin{figure}
    \centering
    \includegraphics{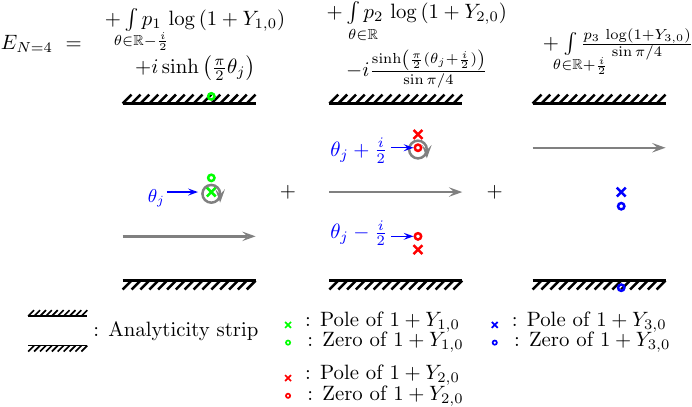}
    \caption{Analyticity of the integrand for the Energy and choice of
      the contours for \(N=4\) }
    \label{fig:U1_contour4}
  \end{figure}
\end{DIFnomarkup}

When \(N\) is even, the corresponding contour cannot be chosen as a 
straight line. We will conjecture here  the analogue of \eqref{eq:oddNU1energy} to be 
simply 
\begin{eqnarray} \label{eq:eveNU1energy}       
E(\L) &=& -\frac{m}{N} \sum_{a = 1}^{N-1} \int_{-\infty-N\iq+a\id}^{\infty-N\iq+a\id}        
\tp_a(\th) \log \left( 1 +       
 Y_{a,0}(\th)        
\right)\mathrm{d}\th\,\,
+\sum_j \cosh(\theta_j)
\end{eqnarray} 
where the roots \(\th_j\) are defined by \(T_{\frac N 2,0}(\th_j)=0\),
so 
that the second term in  
\eqref{eq:eveNU1energy} is real due to the reality of \(T_{\frac N 
  2,0}\). The corresponding contour is shown in figure \ref{fig:U1_contour4}. 

We should admit here that this formula for the masses at even \(N\) 
has a status of a natural conjecture. We have not enough of numerical, 
or analytic evidence to be 100\% sure in it. It would be good to 
verify it at least for the mass gap at \(N=4\), numerically and  by 
means of the L\"uscher corrections at large \(\L\).

\section{ Wronskian solution for Hirota equation equivalent  to Y-system }

For the principal chiral field, \(T_{a,s}\) is defined for   
\(a=0,1,\ldots,N\), while \(Y_{a,s}\) is defined for   
\(a=1,2,\ldots,N-1\).   
We can solve the Hirota finite difference  equation \eqref{eq:Hirota} (and the corresponding Y-system) with the appropriate boundary conditions using its  integrability. Any solution of (\ref{eq:Hirota}) is gauge equivalent to a solution   
where \(T_{0,s}(\th)=T_{0,0}(\th-s\id)\) and   
\(T_{N,s}(\th)=T_{N,0}(\th+s\id)\). We will chose this convention for
the gauge \(T^{(R)}\):
\begin{align}\label{eq:6}
  T_{0,s}^{(R)}(\th)=&T_{0,0}^{(R)}(\th-s\id)\,,&\textrm{and}&&
T_{N,s}^{(R)}(\th)=&T_{N,0}^{(R)}(\th+s\id)\,.
\end{align}

 The most general solution under this gauge constraint
can be expressed \cite{Krichever:1996qd} as an \(N\times N\) determinant,   
in terms of \(2 N\) unknown functions \(q_j\) and   
\(\overline{q_j}\)~\footnote{The general solution doesn't assume that   
  \(q\)'s and \(\overline{q}\)'s are complex-conjugated.  
  Nonetheless, our numerics has shown that at least for the states in  
  the \(U(1)\) sector, it is sufficient to restrict ourselves to the
  solutions where    \(q\)'s and \(\overline{q}\)'s are complex
  conjugated.}\(^{,}\)\footnote{The present q-functions \(q_i\) are related to
  the functions  \(Q_{i}\) in the Hasse diagram
  notation of \cite{Tsuboi:2009ud}.}:    
\begin{gather}   
\label{eq:DetDef}  
  T_{a,s}^{(R)}(\th)=i^{\frac {N(N-1)}2}\mathrm{Det}(c_{j,k})_{1\leq j,k\leq N}   \\
  \begin{aligned}
    \textrm{where\qquad}c_{j,k}=&\overline{q_j}\left(\th+\id\left(s+a+1+\frac
        N 2 -2 k\right)\right)\textrm{\qquad if \(k\leq a\)}\nonumber\\
    \textrm{and\qquad}c_{j,k}=&q_j\left(\th+\id\left(-s+a+1+\frac N 2
        -2 k\right)\right)\textrm{\qquad if \(k> a\)}\,.\nonumber
  \end{aligned}
\end{gather}          
At this point, \(q_j\) is not necessarily the complex-conjugate of   
\(\overline{q_j}\) and the gauge freedom reduces to two independent   
functions \(g\) and \(\overline g\)   
\begin{eqnarray}   
\label{eq:gaug1}
  q_j(\th)&\to&g(\th)\cdot q_j(\th)\\   
  \overline{q_j}(\th)&\to&\overline{g}(\th)\cdot \overline{q_j}(\th)   
\label{eq:gaug2}
\end{eqnarray}        
As an example of this determinant solution, the large \(\L\) (spin
chain limit) solution corresponding to the states of \(U(1)\) sector,
described by the  roots \(\th_i\), can be easily identified %
by plugging the following values for \(q_j\) into   
(\ref{eq:DetDef})   
\begin{gather}   
\begin{aligned}
\label{eq:U1SpChF}
    q_j(\th)=&\overline{q_j}(\th)=\frac {\th^{j-1}}{(j-1) !} &&\textrm{ for }1\le j<N\\
    q_N(\th)=&\overline{q_N}(\th)=P_{\infty}(\th)
  \end{aligned}\\
\textrm{where}\;\;\;\;\;(i\,e^{-\frac i   
  2 \partial_\th}-i\,e^{\frac i   
  2 \partial_\th})^{N-1}P_{\infty}=\varphi\;=\;\prod_k(\th-\th_k)   
\label{eq:U1SpChL}   
\end{gather}  
To see that this is the correct parameterization of the \(U(1)\) solution, first, we can convince ourselves that \(T_{a,-1}=0\),   \(T_{a,0}=\varphi(\th-i \frac {N-2a}   
4)\), and second,   
that it reproduces the \(T_{1,s}\) generated by  
(\ref{eq:defW}-\ref{eq:DefMon}) where all \(\left.Q_j(\th)\right|_{j<N}\)  
are set to \(1\).  For the vacuum state \(P_{\infty}(\th)=\frac{\th^{N-1}}{(N-1)!}\).

In the gauge \(T^{(L)}\), by contrast, we symmetrically chose \(T_{0,s}^{(L)}(\th)=T_{0,0}^{(L)}(\th+s\id)\) and   
\(T_{N,s}^{(L)}(\th)=T_{N,0}^{(L)}(\th-s\id)\), and we have
\begin{gather}   
  T_{a,s}^{(L)}(\th)=i^{\frac{N(N-1)}2}\mathrm{Det}(c_{j,k}')_{1\leq j,k\leq N}   \\
  \begin{aligned}
    \textrm{where\qquad}c_{j,k}'=&\overline{q_j}'\left(\th+\id\left(-s+a+1+\frac
        N 2 -2 k\right)\right)\textrm{\qquad if \(k\leq a\)}\nonumber\\
    \textrm{and\qquad}c_{j,k}'=&q'_j\left(\th+\id\left(s+a+1+\frac N 2
        -2 k\right)\right)\textrm{\qquad if \(k> a\)}\,.\nonumber
  \end{aligned}
\end{gather}

Now we will explain how this allows to generalize the large \(\L\)
solution of section \ref{sec:SpinChain} to any finite \(\L\).

\section{ Solution of the \texorpdfstring{$Y$}{Y}-system for PCF at a finite volume \texorpdfstring{$L$}{L} }   
\label{sec:algo}  

This section describes how to solve the finite volume \(Y\)-system by reducing it to a finite   
number of non-linear integral equations (NLIEs), that can be solved  in its turn  by iterative numerical   
methods.   

We will focus on \(U(1)\) sector states, although the method is in   
principle applicable to any excited state (see the discussion in   
subsection \ref{general_excited}).

\subsection{Definition of the jump densities}   
\label{sec:q-ansatz}

We propose here an ansatz for the finite size \(\L\) solution  by 
adding to the large   
\(\L\)   polynomial expressions (\ref{eq:U1SpChF}-\ref{eq:U1SpChL}) for   
\(q\)'s certain  terms decreasing for \(\theta\to\pm\infty\) and 
exponentially small for    
\(\L\to\infty\) or \(\th\to\infty\) :   
The finite \(\L\)   \(q_j\)'s take thus the form   
\begin{eqnarray}        
  \label{eq:qFinitSizeU1F}        
  q_j(\th)&=&\frac{\th^{j-1}}{j-1 !}+F_j(\th)\qquad \textrm{When        
  }j<N\textrm{ and }\im (\th)\leq 0\\        
  \overline{q_j}(\th)&=&\frac{\th^{j-1}}{j-1 !}+\overline{F_j}(\th)        
  \qquad \textrm{When }j<N\textrm{ and }\im (\th)\geq0\\        
  q_N(\th)&=&P(\th)+F_N(\th)\qquad \textrm{When        
  }\im (\th)\leq 0\\        
  \overline{q_N}(\th)&=&P(\th)+\overline{F_N}(\th)         
  \qquad \textrm{When }\im (\th)\geq0        
\label{eq:qFinitSizeU1L}   
\end{eqnarray}   
where         
\begin{eqnarray}        
  \label{eq:FDef}        
  F_j(\th)&=&\frac 1{2i \pi} \int_{-\infty}^\infty 
  \frac{f_j(\eta)}{\th-\eta}\mathrm{d}\eta \qquad \textrm{When 
  }\im (\th)< 0\\        
\label{eq:FbDef} 
  \overline{F_j}(\th)&=&\frac 1{2i \pi} \int_{-\infty}^\infty \frac{f_j(\eta)}{\th-\eta}\mathrm{d}\eta \qquad \textrm{When }\im (\th)> 0        
\end{eqnarray}        
and the polynomial \(P\) has the same degree as 
\(P_{\infty}=\lim_{\L\to\infty}P\) given by 
(\ref{eq:U1SpChL})\footnote{The way we fix this polynomial will 
 be explained in section \ref {sec:FSBE}, where the finite
   size Bethe equations are discussed.}. Note that for the vacuum
 state at finite \(L\) we have to choose\footnote{For vacuum, it would
 in principle be possible to set \(P\) to any polynomial of degree
 \(N-1\). However, terms of lower degree than \(N-1\) can be set to zero
 by operations on lines and columns of the determinant
 \eqref{eq:DetDef}, and the normalization can be fixed to
 \(P=\frac{\th^{N-1}}{(N-1)!}\) at the price of changing the
 normalization of \(T\)-functions (a particular case of gauge
 transformation).} again \(P=\frac{\th^{N-1}}{(N-1)!}\). As a  consequence of these
definitions, we have
\begin{equation}
\label{eq:jump}
  \overline{q_j}^{[+0]}-q_j^{[-0]}\equiv \lim_{\epsilon\to 0}\overline{q_j}^{[+\epsilon]}-q_j^{[-\epsilon]}=-f_j
\end{equation}
so that \(f_j\) is actually the discontinuity (jump) between the functions
  \(q_j\) and \(\overline{q_j}\) on the real axis.

 Eqs.(\ref{eq:qFinitSizeU1F}-\ref{eq:qFinitSizeU1L})  define   
\(q_j\) only below the real axis and \(\overline{q_j}\) above   
the real axis, so that the determinant only allows to compute        
\(T_{a,s}\) inside the strip \(\im (\th)\in [- \frac {N-2a}        
4-\frac{s+1}  2,-\frac {N-2a} 4+\frac{s+1}  2]\). We can already see 
that these strips are the minimal strips to compute the \(Y\) functions on the 
integration contour of equation \eqref{eq:EnergyOdd}, and we will see 
that it enables to compute the exact\footnote{up to the precision of 
  our numerical         
  procedure solving these NLIE's.} energy of states in the \(U(1)\) 
sector, at any length \(\L\).

The jump densities \(f_j(\eta)\) are well defined on the real axis  
where they take only imaginary values, which follows from \eqref{eq:jump}
 and they  are  exponentially suppressed at  large \(\L\) or large \(\cosh(\frac{2\pi}{N}\theta)\), as can be inferred from \eqref{eq:expSupr}. 

At the end of this subsection, let us comment on a slightly
generalized version of our Wronskian solution of this section, now
including the twisted, quasi-periodic boundary conditions on the wave
function of the system. The twist matrix \(g\in SU(N)\) can be chosen,
without loss of generality, in a diagonal form: \(g={\rm
  diag}\{x_1,x_2,\cdots,x_N\}\) where the eigenvalues are  unitary: \(\bar x_j=\frac{1}{x_j},\) and \(\prod_{j=1}^Nx_j=1\). Then the ansatz
\eqref{eq:qFinitSizeU1F}-\eqref{eq:qFinitSizeU1L} will be
modified\footnote{In this Ansatz, one can either chose to write
  \(e^{F_j}\) or \(1+F_j\), it only amounts to a slight change of the
  functions \(F_j\). What really differs with respect to
  \eqref{eq:qFinitSizeU1F}-\eqref{eq:qFinitSizeU1L} is the factor
  \(x_j^{i\th}\) and the fact that the polynomial term
  \(\frac{\th^{j-1}}{j-1 !}\) is replaced by \(1\).} \begin{eqnarray}        
  \label{eq:qFinitSizeU1Ftw}        
  q_j(\th)&=&x_{j}^{i\th}e^{F_j(\th)}\qquad \textrm{When        
  }1 \le j<N\textrm{ and }\im (\th)\leq 0\\        
  \overline{q_j}(\th)&=&x_{j}^{i\th}e^{\overline{F_j}(\th)}        
  \qquad \textrm{When        
  }1 \le j<N\textrm{ and }\im (\th)\geq0\\        
  q_N(\th)&=&P(\th)x_{N}^{i\th}e^{F_N(\th)}\qquad \textrm{When        
  }\im (\th)\leq 0\\        
  \overline{q_N}(\th)&=&P(\th)x_{N}^{i\th}e^{\overline{F_N}(\th)}         
  \qquad \textrm{When }\im (\th)\geq0        
\label{eq:qFinitSizeU1Ltw}   
\end{eqnarray}      with the same definition  \eqref{eq:FbDef} for the
functions \(\bar F,F\) (we can put \(F_1=\overline{F_1}=0) \). For the
vacuum state, we should put \(P(\th)=1\). In this case,  in the limit
\(L\to\infty\) we obtain from the formula \eqref{eq:DetDef} that the
T-function becomes a character of representation \(\lambda=a^s\) of
the twist matrix (up to the Vandermonde determinant
\(\Delta(x_1,\cdots,x_N)\)).  Note that it is not a trivial matter to
reproduce in the untwisting  limit \(x_j\to 1,\quad j=1,2,\dots,N\)
the  ansatz \eqref{eq:qFinitSizeU1F}-\eqref{eq:qFinitSizeU1L}: One has
to do special rotations of the basis of \(q_j\) to arrive at the right
answer.

The present article describes symmetric states (more specifically
\(U(1)\) sector states), for which there is a single twist matrix \(g\in
SU(N)\). But in general, there should be two independent \(SU(N)\)
twists due to the overall \(SU(N)\times SU(N)\) symmetry.
If we use a general \(SU(N)\times SU(N)\) twist and break the symmetry
of the state, then one
would need two distinct sets of q-functions for the right and left
wing, as argued in the discussion about non-symmetric states (in
section \ref{general_excited}). In that case, one twist appears in the
parameterization (\ref{eq:qFinitSizeU1F}-\ref{eq:qFinitSizeU1Ltw}) of
the right wing, and another twist in the parameterization of the left wing.

\subsection{Relation to the analyticity of \texorpdfstring{$T$}{T} functions} 
From the ABA \eqref{eq:largeLY1} and the finite size equation  
\eqref{eq:YcentralR}, we can see that    
\begin{eqnarray}        
  1+Y_{a,0}&\xrightarrow[\substack{\th \to\infty\\\textrm{or }L\to\infty}]{}&1\qquad \textrm{when        
  }|\im (\th)|<\frac N 4        
\end{eqnarray}         
which means that \(1+Y_{a,0}=\frac{T_{a,0}^+T_{a,0}^-}  
{T_{a+1,0}T_{a-1,0}}\) has a proper behavior in this strip, being a  
meromorphic function regular at infinity.  
On the other hand, when \(|\im (\th)|=\frac N 4\), \(1+Y_{a,0}\) oscillates  
at \(Re(\th)\to\infty\), and it diverges when, e.g., \(|\im (\th)|\in[\frac N  
4,\frac {3 N} 4]\).  
By that reason we conclude that the analyticity\footnote{Note that we
  use the word ``analyticity strip'' to denote a %
  domain where
  the \(L\to\infty\) limit of Y- (resp T- and q-)functions is a
  meromorphic (resp holomorphic) function of \(\theta\).} strip of  
\(1+Y_{a,0}=\frac{T_{a,0}^+T_{a,0}^-} {T_{a+1,0}T_{a-1,0}}\) is \(\{\th  
,|\im (\th)|<\frac N 4\}\). From this we can identify the strips where
the asymptotics \eqref{eq:phiTboundRi} hold for \(T_{a,s}\):

\begin{eqnarray}        
\label{eq:stripsforTf} 
  T_{0,0}&\xrightarrow[\L \cosh(\frac {2\pi\th} N)\to\infty]{}&\varphi^{[-N/2]}\qquad \textrm{when        
  }\im (\th)<\frac N 4\\        
  \left.T_{a,0}\right|_{0<a<N}&\xrightarrow[\L \cosh(\frac {2\pi\th} N)\to\infty]{}&\varphi^{[+a-N/2]}\qquad \textrm{when        
  }|\im (\th)|<\frac N 4+\frac 1 2\\        
  T_{N,0}&\xrightarrow[\L \cosh(\frac {2\pi\th} N)\to\infty]{}&\varphi^{[+N/2]}\qquad \textrm{ when        
  }\im (\th)>-\frac N 4        \,.
\label{eq:stripsforTl} 
\end{eqnarray}         
  These conditions ensure the proper analyticity of  
  \(1+Y_{a,0}=\frac{T_{a,0}^+T_{a,0}^-} {T_{a+1,0}T_{a-1,0}}\), and  
  the boundaries of the analyticity strips of each \(1+Y_{a,0}\) are  
  given by the boundaries of the analyticity strips of the  
  corresponding \(T\) functions\footnote{The analyticity strips for  
    \(T_{0,0}\) and \(T_{N,0}\) can be chosen on a half plane thanks to an  
    appropriate gauge.}.  

Now, since we know that the \(T\) functions are described by Wronskian 
determinants, these analyticity strips suggest that 
\begin{equation} 
\label{eq:anforq} 
  q_j\qquad\textrm{is analytic 
when}\qquad\im (\th)<1/2 
\end{equation} 
\begin{equation} 
\label{eq:anforbq} 
  \overline{q_j}\qquad\textrm{is analytic 
when}\qquad\im (\th)>-1/2 
\end{equation} 
So the analyticity strip 
for \(q_j\) 
ends up at \(\im (\th)=1/2\) which is reflected for instance in the 
fact that \(T_{N-1,0}\) is not analytic when \(\im (\th)>N/4+1/2\). 
This explains why \(Y_{N-1,0}\) isn't analytic when \(\im (\th)>N/4\). 

The equations (\ref{eq:anforq},\ref{eq:anforbq}) teach us
  that the analyticity domain is a bit bigger than what is necessary for
  (\ref{eq:qFinitSizeU1F}-\ref{eq:qFinitSizeU1L}). It tells us
that in the definitions (\ref{eq:FDef},\ref{eq:FbDef}), the contour 
can be shifted up to \(\pm i/2\). In other words, the functions \(f_j(\eta)\) are 
analytic on the strip \(|\im (\eta)|<1/2\).

It is noteworthy that even with these contour deformations, the
determinant expressions (\ref{eq:qFinitSizeU1F}-\ref{eq:qFinitSizeU1L}) describe 
the function \(T_{a,s}(\th)\) inside the strip \(\im (\th)\in ]- \frac {N-2a}        
4-\frac{s}  2-1,-\frac {N-2a} 4+\frac{s}  2+1[\), which is narrower 
than in equations (\ref{eq:stripsforTf},\ref{eq:stripsforTl}). But we 
will show that the relatively narrow strips given by this ansatz are 
sufficient to solve the \(Y\)-system and compute the energies. 

\subsection{Closed system of  NLIEs}  
   
The gauge freedom  (\ref{eq:gaug1},\ref{eq:gaug2})
can be used to impose  \(F_1(\th)=\overline{F_1}(\th)=0\), which leaves        
only \(N-1\) independent densities to compute.        
Let us now see how \(N-1\) equations on this densities can be obtained by imposing that the        
state is symmetric, i.e. that \(Y_{a,-s}=Y_{a,+s}\). This requirement
means that we can chose \(T^{(L)}_{a,-s}=T_{a,s}\), which simplifies %
the \(Y\)-system equation  for the  middle node        
(\ref{eq:Ycentral})\footnote{As a consequence, 
we are solving the \(Y\)-system under the two following constraints : 
\(Y_{a,s}\sim e ^ {-\L \tp_a (\theta) \delta_{s,0}}\times 
\mathrm{const}_{a,s}\) on the one hand, and  
\(T^{(L)}_{a,-1}=T_{a,1}\)  one the other hand. This second 
constraint is specific to symmetric states (which includes the states 
in the \(U(1)\) sector), such that \(Y_{a,-s}=Y_{a,s}\).}, in the same
manner as in \cite{Gromov:2008gj}.
    
Using this symmetry of the state and the boundary condition \eqref{eq:6}, the
equation (\ref{eq:YcentralR}) is reduced to:         
\begin{align}        
\label{eq:YeqFC}        
    Y_{a,0}=& e^{-\L \tp_a}  \frac{(T_{a,1})^2}        
      {T_{a-1,0}T_{a+1,0}}
\left(\frac{T_{0,0}^{[+N-a]}T_{N,0}^{[-a]}}{T_{0,0}^{[a-N]}T_{N,0}^{[+a]}}\right)^{\star
  K_N}        \,,
\intertext{or equivalently,}
    T_{a,-1}^{[a-N/2]}=& e^{-\L \tp_a^{[a-N/2]}}  T_{a,1}^{[a-N/2]}
\left(\frac{T_{0,0}^{[+N/2]}T_{N,0}^{[-N/2]}}{T_{0,0}^{[2a-3N/2]}T_{N,0}^{[2a-N/2]}}\right)^{\star
  K_N}        \,.\label{eq:4}
\end{align}

The reason why we chose such shifts in the relation \eqref{eq:4} is
that the l.h.s. has a determinant expression  
(\ref{eq:DetDef}) %
where one has
\(c_{j,a}=\overline{q_j}^{[+0]}\)  while    
\(c_{j,a+1}={q_j}^{[-0]}\). After  subtracting one column from  
another in the determinant, there is a full column of    
\(\overline{q_j}^{[+0]}-q_j^{[-0]}=-f_j\) which is exponentially small. 
That explains    
the exponential suppression of \(T_{a,-1}\). Expanding the determinant
w.r.t. these columns\footnote{The two terms in \eqref{eq:LinSysC}
  correspond to the fact that before expanding the determinant, we have added
  and subtracted columns to get one full column of
\(\overline{q_j}^{[+0]}-q_j^{[-0]}=-f_j\) and the other one of
\(\frac{\overline{q_j}^{[+0]}+q_j^{[-0]}}2\) (which corresponds to the principal
value  in (\ref{eq:FDef},\ref{eq:FbDef})).} gives the following linear
system relating \(f_j\)'s to \(T_{a,-1}\)'s:   
\begin{gather}
  \label{eq:LinSysQ}   
  T_{a,-1}\left(\th-i \frac {N-2a} 4\right)=\sum_j d_{a,j}(\th)
  f_j(\th)\\   
  \label{eq:LinSysC}   
\textrm{where}\;\;d_{a,j}=i^{\frac{N(N-1)}{2}}(-1)^{j+a+1}\frac{
    \det(c_{k,l})_{\substack{k\neq   
      j\\l\neq a}}+ \det(c_{k,l})_{\substack{k\neq j\\l\neq   
      a+1}}
}2   
\end{gather}
These \(c_{k,l}\) are the coefficients of the   
determinant (\ref{eq:DetDef}) defining \(T_{a,-1}\left(\th-i \frac   
  {N-2a} 4\right)\), and finally   
equations (\ref{eq:YeqFC},\ref{eq:LinSysQ}) can be recast into     
\begin{eqnarray}   
\label{eq:closed}   
\sum_j d_{a,j}(\th) f_j(\th)&=& e^{-\L \tp_a(\th-i \frac   
  {N-2a} 4)}
      T_{a,1}^{[a-N/2]}
\left(\frac{T_{0,0}^{[+N/2]}T_{N,0}^{[-N/2]}}{T_{0,0}^{[2a-3N/2]}T_{N,0}^{[-N/2+2a]}}\right)^{\star         
  K_N}   \,.
\end{eqnarray}  
This  is a closed system of equations on  
\(\{f_j(\th)\}_{\th\in \mathbb{R}}\)  
because all coefficients    
\(d_{a,j}\), and all \(T\)'s can be computed out of \(f_j\)'s through several   
convolutions.   

The solution of the \(Y\)-system is therefore achieved by solving this   
system of \(N-1\) equations on \(N-1\) densities. The simple   
inversion of the linear system (\ref{eq:LinSysQ}) brings   
(\ref{eq:closed}) into the form   
\begin{equation}   
  \label{eq:CloSim}   
  f_j(\th)=H_j(\{f_k(\eta)\}_{\substack{k=2\ldots N-1\\\eta\in\mathrm{R}}})  \,.
\end{equation}   
This \(H_j\) defines a contraction mapping in some 
vicinity of \(f_j=0\) when \(\L\) is    
sufficiently large since it leads to an exponentially
small \(f_j(\th)\).  This implies that in some vicinity of 
\(\L=\infty\), the mapping \(H_j\) has a   
fixed  point that can be  found numerically through repeated iterations  
of  \(H\).   

The way we solve \(Y\)-system is therefore simply the iteration of
\eqref{eq:CloSim}  and a good news is that, at least for \(N=3\), even
at very small \(\L\), this    
procedure seams, at least according to our numerics, to converge to a fix point of \(H_j\), giving a
complete solution    
of \eqref{eq:CloSim} and thus of the \(Y\)-system.

\subsubsection{Numerically workable form for the NLIE's}

One difficulty of this numerical process is  in        
computing         
\(\left(\frac{T_{0,0}^{[+N/2]}T_{N,0}^{[-N/2]}}{T_{0,0}^{[2a-3N/2]}T_{N,0}^{[-N/2+2a]}}\right)^{\star         
  K_N}\) in the  right hand side of \eqref{eq:closed}. As we have already seen,         
the Wronskian formula \eqref{eq:DetDef} in terms of \(q_j\)'s having cuts on the real axis  
allows to compute \(T_{0,0}(\th)\) only         
when \(\im (\th)<-\frac N 4+\frac 1 2\). %
So, for instance,   
\(T_{0,0}(\th+i\frac N 4)\) cannot be computed in this way when the spectral parameter  
\(\th\) is real. The denominator can        
nonetheless be computed if  in the convolutions we shift appropriately  both the argument of the kernel and of the T-functions        
\begin{gather}
\label{eq:shimove}  
{  \left(\frac{1}{T_{0,0}^{[2a-3N/2]}T_{N,0}^{[-N/2+2a]}}\right)^{\star         
  K_N}}%
= \left(\frac{1}{T_{0,0}^{[-N/2-N+a + 1]}}\right)^{\star         
  {K_N}^{[a - 1]}}\left(\frac{1}{T_{N,0}^{[+N/2+a - 1]}}\right)^{\star         
  {K_N}^{[-N+a + 1]}}        
\end{gather}
since \(K_N(\th)\) is regular when \(\im (\th)\in[-\frac  
{N-1}2,\frac {N-1}2]\) and   \(a\in[1,N-1]\).  
Eq.\eqref{eq:shimove} simply reflects the fact that if \(k(\th)\) is analytic  
for \(\im (\th)\in[0,b/2]\), then  for real \(\th\),  
\(\left(f^{[b]}\right)^{\star   k}=f^{\star  (k^{[b]})}\,,\;\; \forall f\). %

The same idea, applied to the numerator, would give   
\(\left(T_{0,0}(\th+N \iq)\right)^{\star  K_N}=\left(T_{0,0}(\th-N  
  \iq)\right)^{\star  K_N^{[+N]}}\). But the equality fails because   
\({K_N}\) has a pole at \(i \frac {N-1}2\). %

Instead, one can use the following relation        
\begin{eqnarray}        
\label{eq:mytrick}        
   \left(T_{0,0}\left(\th+N \iq\right)\right)^{\star  K_N}         
&=&\frac{T_{0,0}(\th-N \iq+\id)}{\left(T_{0,0}^{[+\frac N 2 -2]} T_{0,0}^{[+\frac N 2-4]}\cdots        
      T_{0,0}^{[-3 \frac N 2+2]} \right)        
^{\star  K_N}}\\        
\label{eq:mysectrick}        
&=&\frac{T_{0,0}(\th-N \iq+\id)}{T_{0,0}(\th-N \iq-\id)}  
\left(T_{0,0}^{[-3N/2]}\right)^{\star K_N}  
\end{eqnarray}        
which simply uses the fact that for a regular function  \(f\), \(   
\left(\MyPi_{N}[f]\right)^{\star K_N}=f\). This is true only up to a zero mode of  
{\(\MyPi_N\)} which will be discussed in the section \ref{sec:chiCDD}.
    
Finally, the last factor in the eq.\eqref{eq:closed} can be put into a
numerically workable form        
by rewriting        
\begin{multline}
\label{eq:trickeq}        
{  
  \left(\frac{T_{0,0}^{[+N/2]}T_{N,0}^{[-N/2]}}{T_{0,0}^{[2a-3N/2]}T_{N,0}^{[-N/2+2a]}}\right)^{\star         
  K_N}=}%
\\        
\frac{T_{0,0}^{[-N/2+1]}}{T_{0,0}^{[-N/2-1]}}  
\frac{T_{N,0}^{[N/2-1]}}{T_{N,0}^{[N/2+1]}}   
\left(  
\frac{T_{0,0}^{[-3N/2-a+1]}}{T_{0,0}^{[-3N/2+a%
{+}1]}}  
\right)^{\star K_N^{[a-1]}}  
\left(  
\frac{T_{N,0}^{[3N/2+N-a%
{-}1]}}{T_{N,0}^{[3N/2-N+a-1]}}  
\right)^{\star K_N^{[-N+a+1]}}  
\end{multline}

This will help us to transforms the eq.\eqref{eq:closed}, once the appropriate zero mode is added, into a really  closed system of NLIEs  
where the right hand side can indeed be computed by knowing  the  
functions \(f_j\) only on the real axis.

\subsubsection{\texorpdfstring{$\chi_{_{CDD}}$}{chi\_CDD} factor}   

\label{sec:chiCDD}       

It is clear from the derivation of  eq.\eqref{eq:closed}, as well as  
of the eqs.(\ref{eq:mytrick}, \ref{eq:trickeq})  that they  
 are fixed only up to a zero mode of the operator %
{\(\MyPi_N\)}.
A zero mode \(Z\) therefore has to be added to  
\eqref{eq:closed}, to get  
\begin{multline}  
\label{eq:Main}  
  \sum_j d_{a,j}(\th) f_j(\th)= Z~e^{-\L \cosh(\frac{2 \pi} N (\th-i \frac   
  {N-2a} 4))   \frac{
  \sin(\frac{a \pi}N)}{
  \sin(\frac{\pi}N)}}  
      T_{a,1}^{[a-N/2]}
      \\          
 \times        
\frac{T_{0,0}^{[-N/2+1]}}{T_{0,0}^{[-N/2-1]}}  
\frac{T_{N,0}^{[N/2-1]}}{T_{N,0}^{[N/2+1]}}   
\left(  
\frac{T_{0,0}^{[-3N/2-a+1]}}{T_{0,0}^{[-3N/2+a%
{+}1]}} 
\right)^{*K_N^{[a-1]}}  
\left(  
\frac{T_{N,0}^{[3N/2+N-a%
{-}1]}}{T_{N,0}^{[3N/2-N+a-1]}}  
\right)^{*K_N^{[-N+a+1]}}  %
\end{multline}
 Such zero  
mode can include for instance the factors \(e^{-\L \cosh(\frac{2 \pi \th} N )   
      }\) and  \(\chi_{_{CDD}}\) from equation
 \eqref{eq:3}.

In the asymptotic limit (\(\L\to \infty\)), the zero modes in the equation  
(\ref{eq:YeqFC}) can be obtained by comparison with  
\eqref{eq:largeLY1}.  
In the same manner, the zero modes implicitly present in equation  
\eqref{eq:trickeq} can be computed in the asymptotic limit, by replacing  \(T_{a,0}\)'s  by their asymptotic values in terms of    
\(\varphi\), see   eq.\eqref{eq:phiTboundRi}, so that we can
directly compute the zero mode \(Z\) when \(\L\to\infty\).
We notice that, although both  
\eqref{eq:closed} and \eqref{eq:trickeq} are true up to a non-trivial  
zero mode, the zero mode in \eqref{eq:Main} happens to  
be\footnote{\label{fn:1}Actually the right hand side of \eqref{eq:Main} is itself  
  defined up to a    
  factor of \(e^{i \frac{2 k \pi}{N}}\), because any \(f^{\star  K}=e^{K*\log  
      f}\) is defined up to a \(e^{2 i \pi \int K}\) corresponding to the  
  choice of the branch of the log. As a consequence, a more precise  
  statement for \eqref{eq:Z1} is \(Z=e^{i \frac{2 k \pi}{N}}\).  \(k\) is chosen to reproduce \eqref{eq:largeLY1}, where the phase  
  in \eqref{eq:Smatr} is chosen in such a way that one particle at  
rest (\(\theta_1=0\)) is a solution of the Bethe equation  
\eqref{eq:BAE}.  

For states with zero momentum (like the mass gap and the vacuum), this extra  
phase can also be obtained by requiring a \(\th\to -\th\) symmetry, which  
these states should exhibit.  
}  
\begin{eqnarray}  
  \label{eq:Z1}  
  Z=1,  
\end{eqnarray}  
at least in the asymptotic limit!

Indeed, if we compare \eqref{eq:Main} with \eqref{eq:largeLY1} (where
\(Y_{a,0}=\frac {T_{a,1}T_{a,-1}}{T_{a+1,1}T_{a-1,-1}}\) is expressed
using \eqref{eq:LinSysQ}), we see that in the asymptotic limit we have
\begin{multline}
  \label{eq:1}
Z    
\frac{T_{0,0}^{[-N/2+1]}}{T_{0,0}^{[-N/2-1]}}  
\frac{T_{N,0}^{[N/2-1]}}{T_{N,0}^{[N/2+1]}}   
\left(  
\frac{T_{0,0}^{[-3N/2-a+1]}}{T_{0,0}^{[-3N/2+a%
{+}1]}} 
\right)^{*K_N^{[a-1]}}  
\left(  
\frac{T_{N,0}^{[3N/2+N-a%
{-}1]}}{T_{N,0}^{[3N/2-N+a-1]}}  
\right)^{*K_N^{[-N+a+1]}} 
\\
\sim  
\frac{\varphi^{[-N+1]}}{\varphi^{[-N+2a-1]}}   
\frac{\varphi^{[-N+1]}}{\varphi^{[-N+2a+1]}}   
\frac 1 {\MyPi_{a}\left[\left(S^{[a-N]}\right)^2\chi_{_{CDD}}^{[a-N]}\right]}  \,,
\end{multline}
from where we will show that \(Z_{\infty}=1\), where \(Z_{\infty}\) denotes
the asymptotic limit of \(Z\), which  can be extracted from
\eqref{eq:1}. To this end, we can note that, as a 
direct consequence of (\ref{phi-function},\ref{eq:2},\ref{eq:3},
\ref{eq:stripsforTf}-\ref{eq:stripsforTl}), we have
\(Z_{\infty}(\th)=\prod_j Z_0(\th-\th_j)\), where \(Z_0\) is the value of
\(Z_\infty\) corresponding to \(\varphi(\th)=\th\) (i.e. one single root at
the origin). It is therefore sufficient to show that \(Z_0=1\), i.e. to
study the equation \eqref{eq:1} when \(\varphi(\th)=\th\). In this case, we can easily list the zeroes and poles of the r.h.s of \eqref{eq:1}:  %
its zeroes are at positions
\(%
\frac {N-1}2+k\,N\), \(%
\frac{3N+1}2 -a+k\,N\), \(%
-
\frac{N-1}2-k\,N\) and
\(%
-
\frac{N+1}2-a-k\,N\) (where \(k\geq 0\) is an arbitrary non-negative
integer) and its poles are at positions \(%
\frac {N+1}2+k\,N\),
\(%
\frac{3N-1}2 -a+k\,N\), \(%
-\frac{N+1}2-k\,N\) and \(%
-\frac{N-1}2-a-k\,N\).

We can then substitute the asymptotic limit
(\ref{eq:stripsforTf}-\ref{eq:stripsforTl})  into the l.h.s to find
its analytic properties: one sees that the factor \(\frac{T_{0,0}^{[-N/2+1]}}{T_{0,0}^{[-N/2-1]}}  
\frac{T_{N,0}^{[N/2-1]}}{T_{N,0}^{[N/2+1]}}   
\) reproduces the same zeros and poles as the r.h.s. at position \(\pm
\frac {N\pm 1} 2\), whereas the factors \((\dots)^{\star K_N}\) %
are
analytic when \(-\frac {N-1} 2 -a<\im (u)<\frac{N-1} 2 + N-a\). 
Hence we see that in the asymptotic limit, \(Z\) is analytic in the
strip \(-\frac {N-1} 2 -a<\im (u)<\frac{N-1} 2 + N-a\), but as a zero-mode it
is also \((N\,i)\)-periodic. Moreover it behaves as a constant when
\(\mathrm{Re}(u)\to\infty\), so that Liouville theorem implies that \(Z_0\)
is a constant, hence (as a zero mode) it is equal to \(e^{i \frac{2 k
    \pi}{N}}\) for a given value of \(k\). This factor can be absorbed
into the ambiguity in the definition of the branch of the logarithm in
the definition of 
\(f^{\star  K_N} = e^{\log f \star  K_N}\)
(see also footnote
\ref{fn:1}).

 In the numerical solution of the  
\(Y\)-system we therefore assume that \(Z=1\) holds even at a finite size,  
i.e. that the analyticity structure of  the zero modes is the  
same at finite \(\L\) as at \(\L\to \infty\).  
We explicitly see that at \(L\to\infty\), \(\chi_{_{CDD}}\)
  defined in \eqref{eq:DresPhas} 
   is taken into account in \eqref{eq:Main}. In addition we can check
   that at finite size, we obtain \(Y\) functions having simple poles only.

\subsection{Finite size Bethe equations}   
\label{sec:FSBE} 
Bethe equations emerge in this procedure as a regularity requirement on   
the jump densities \(f_j\)'s. Let us illustrate it for a general \(U(1)\)   
state in the \(SU(3)\) case, and also show  why these  finite \(\L\) analogues of Bethe equations are equivalent, at large \(\L\), to the ABA Bethe equations on the roots   
of \(\varphi\).

For such a state, the linear system (\ref{eq:LinSysQ}) can be written   
as   
\begin{eqnarray}   
  \left(   
    \begin{array}{cc}   
      A&B\\   
      -\overline{A}&-\overline{B}   
    \end{array}   
  \right)   
  \left(   
    \begin{array}{c}   
      f_2\\   
      f_3   
    \end{array}   
  \right)&=&   
  \left(   
    \begin{array}{c}   
      T_{1,-1}(\th-i/4)\\   
      T_{2,-1}(\th+i/4)   
    \end{array}   
  \right)   
\end{eqnarray}   
where   
\(A=\frac i 2 (q_3+\overline{q_3})-i\,q_3^{--}\) and   
\(B=-\frac i 2 (q_2+\overline{q_2})+i\,q_2^{--}\). 

Inverting the matrix \(  \left(   
    \begin{array}{cc}   
      A&B\\   
      -\overline{A}&-\overline{B}   
    \end{array}   
  \right)\), some singularity could occur at the zeroes of the function   
  \(\overline{A}B-A\overline{B}\), i.e. when the determinant is   
  zero. If we want \(f_j\)'s to be regular, we need  the   
  numerator to vanish at the same \(\th\) to cancel this pole. This gives the following finite size   
  Bethe equation:

  \begin{equation}   
\label{eq:finitBE}   
   {\rm If}\quad    
    \left.\left(\overline{A}B-A\overline{B}\right)\right|_{\tth_j}=0\quad{\rm then}\quad       \left\{   
      \begin{array}{rcl}   
T_{1,-1}(\tth_j-i/4)\overline A(\tth_j)&=&-T_{2,-1}(\tth_j+i/4) A(\tth_j)\\   
T_{1,-1}(\tth_j-i/4)\overline B(\tth_j)&=&-T_{2,-1}(\tth_j+i/4) B(\tth_j)   
\end{array}   
\right.   
  \end{equation}   
One can notice that at such  \(\tth_j\)   
the two conditions in the r.h.s.  
 are equivalent.

The claim that \(\tth_j\) are a finite size analogue of the Bethe roots is supported by the fact that at large \(\L\), the roots of   
\(\overline{A}B-A\overline{B}\) are precisely the Bethe roots. Indeed,    
at large \(\L\), \(B\simeq 1\) and \(A\simeq i(P-{P^{--}})\), giving   
\(\overline{A}B-A\overline{B}\simeq i(P^{++}-P+P^{--}-P)= -  
i\varphi\). Moreover, we see from \eqref{eq:7} that
the second  relation  in the r.h.s. of \eqref{eq:finitBE}  
  reduces then to the reality condition   
\(\frac{\overline{T_{1,-1}(\th_j-i/4)}}{T_{1,-1}(\th_j-i/4)}=-1\). Using the   
leading-order large \(\L\) expression of \(Y_{a,0}\) in terms of \(S\), eq.\eqref{eq:largeLY1}, we   
get at large \(\L\)   
\begin{eqnarray}   
  T_{1,-1}(\th-i/4)&\simeq&\frac{\varphi^{--}}{\varphi}\frac{\varphi+2\varphi^{--}}{%
    \mathcal{S}^{--}}   
  e^{-\L\cosh(\frac{2\pi}3 (\th-i/4))}  \\ 
\textrm{where }\mathcal{S}(\th)&=&\prod_jS_0^2(\th-\theta_j){\breve\chi}_{_{CDD}}(\th-\th_j) \,.
\end{eqnarray}   
Using the fact that \(\varphi(\th_i)=0\) at all Bethe roots \(\th_i\), and   
dividing by the complex conjugate, the large \(\L\) regularity   
requirement becomes   
\begin{eqnarray}   
\left.  \frac{\left(\varphi^{--}\right)^2}{\varphi   
    \mathcal{S}^{--}}\frac{\varphi}{\left(\varphi^{++}\right)^2\mathcal{S}^{++}}   
\right|_{\th=\th_i}e^{i \L \sinh(\frac{2\pi}3 \th_j)}&=&-1   \,.
\end{eqnarray}   
Using the crossing relation, the left hand side becomes simply 
\(\mathcal{S}(\th_j)e^{i \L \sinh(\frac{2\pi}3 \th_j)}\), 
so that the  
finite size regularity condition stated above is equivalent at large   
\(\L\) to the asymptotic Bethe equations (\ref{eq:BAE}).   

As a consequence, the iterative solution of the closed, finite size 
equations (\ref{eq:Main}), should start from the expression 
(\ref{eq:qFinitSizeU1F},\ref{eq:qFinitSizeU1L}) where \(P=P_{\infty}\) is 
given in terms of the asymptotic Bethe roots by (\ref{eq:U1SpChL}), 
and then at each iteration, this polynomial is updated in order to 
incorporate this regularity condition.  

\paragraph{Momentum number :}   
{In the asymptotic limit, one can introduce a notion of momentum
number as follows: first one rewrites \eqref{eq:BAE} on the form
\begin{align}
  \forall j&,& e^{i\,L\,\sinh\frac{2\pi}N\theta_j+i\sum_{k\ne j}
    f(\theta_j-\theta_k)}=&1
\end{align}
where the function
\begin{align}
  f(\theta)=&-i \log(-\breve\chi_{_{CDD}}(\theta)S(\theta)^2)
\end{align}
is defined as a monotonous, continuous function, such that
\(f(0)=0\).
This allows to define the mode number of the particle \(j\) as the
integer \(k\) such that \(L\,\sinh\frac{2\pi}N\theta_j+\sum_{k\ne j}
    f(\theta_j-\theta_k)=2\pi k\).

By contrast, at finite size, the regularity condition
(\ref{eq:finitBE}) involves the phase
\(-\frac{T_{1,-1}(\th-i/4)\overline B(\th)}{T_{2,-1}(\th+i/4)
  B(\th)}\). While there are several values of \(\theta\) where this
phase is equal to one, only a few of these values (one for each
particle) are zeroes of \(\overline{A}B-A\overline{B}\); the choice of
these values defines the mode numbers at finite size.

To give a simple example, we can consider a state
with a single Bethe root (\(\mathcal{N}=1\)) and such that
\(L\,\sinh\frac{2\pi}N\theta_0=2\pi\) %
in the asymptotic limit (ie this
    state has momentum number \(1\)). One can easily see that in the
    asymptotic limit the corresponding zero of \(\overline{A}B-A\overline{B}\) is
    the second smallest positive zero of \(\mathrm{Re}(T_{1,-1}(\th-i/4)\overline B(\th))\). Hence, at finite
    size, we  recognize the state with mode number \(1\) as a state
    where the zero of \(\overline{A}B-A\overline{B}\) is
    the second smallest positive zero of
    \(\mathrm{Re}(T_{1,-1}(\th-i/4)\overline B(\th))\). To make sure
    that the iterative resolution algorithm does not ``jump'' from a
    state with given momentum number to another state, the momentum
    number has to be taken into account when the regularity condition
    \eqref{eq:finitBE} is enforced at every iteration.
}

\paragraph{\(N>3\) case :}   
The same construction leads to finite size Bethe equations for any odd  
\(N\).  
 Like in equation (\ref{eq:finitBE}) the number of regularity   
constraints at each zero of the determinant is apparently \(N-1\) but   
reduces to only \emph{one} constraint: the cancellation of the   
projection of   
\(\left(   
    \begin{array}{c}   
      T_{1,-1}(\th-i(N-2)/4)\\   
\vdots\\   
      T_{N-1,-1}(\th+i(N-2)/4)   
    \end{array}   
  \right)\) to the kernel of the matrix \(d_{i,j}\) defining the linear   
  system (\ref{eq:LinSysQ}).   

This procedure for finite-size Bethe equations was described
  here for odd \(N\)  and for states having real Bethe roots in the
  asymptotic limit. The subtlety which arises when \(N\) is even, or for
  the states having, in the asymptotic limit,   complexes
of complex-conjugated Bethe roots, is that the zeroes of the determinant 
do not lie on the real axis but approximately on \(\mathbb{R}\pm
i/2\). The above procedure can in principle be applied anyway, but its
interpretation is left to clarify 
because the regularity condition is imposed at the very boundary of 
the analyticity strip.

\subsection{Numerical results}        
    
\afterpage{\begin{figure}        
  \centering        
\includegraphics{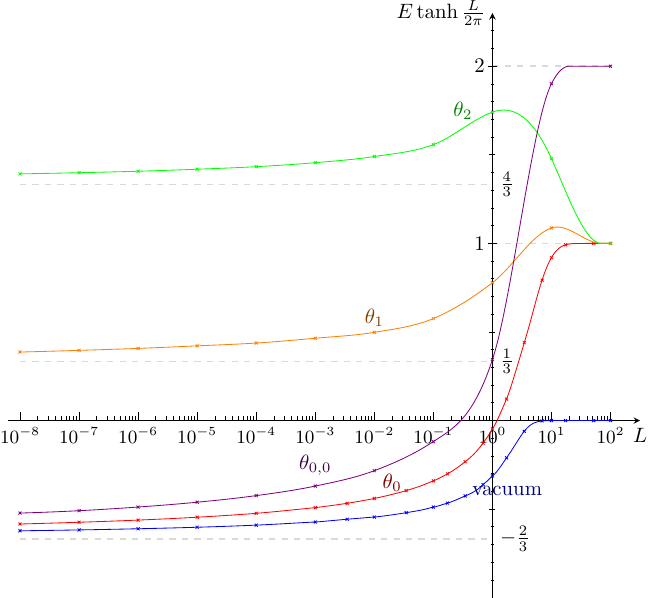}        
  \caption{Energies\protect\footnotemark of  vacuum%
    , of mass gap and of some excited states
    as functions of \(\L\),  at \(N=3\). We see that in the
    \(L\to\infty\) limit, \(E\) tends to the number of ``particles'' (ie
    Bethe roots), whereas
    in the conformal \(L\to 0\) limit, \(E\sim \frac{2\pi}{\L}(-\frac
    {N^2-1}{12}+n)\), where \(n\) is the sum of the mode numbers of the ``particles''.\protect\\
    The curves are interpolations from the numeric points (small crosses).}
  \label{fig:N=3_spec}        
\end{figure}        

\footnotetext{This figure shows the
      combination \(E \tanh{\frac L{2\pi}}\), to make manifest the limit of \(E\) at large \(L\), and of \(E \frac L{2\pi}\) at small \(L\).}

\begin{table} \label{tab:new}
  \centering\( 
\begin{array}{|r|c|c|c|c|c|} 
    \hline%
     \multicolumn{1}{|c|}
    {L}&E_{vacuum}&E_{\th_0}&E_{\th_1}&E_{\th_2}&E_{\th_{0,0}}\\ 
    \hline{} 
    \TP{-8}&-3.90{\gray 9} \tP 8&%
    -3.66{\gray 8} \tP 8 &%
    2.4{\gray 3}
    \tP 8&%
    8.74{\gray 9} \tP 8&%
    -3.{\gray 28} \tP 8\\
    \TP{-7}&%
    -3.878{\gray 0} \tP 7&%
    -3.60{\gray 6} \tP 7 &%
    2.5{\gray 5}
    \tP 7&%
    8.79{\gray 1} \tP 7&%
    -3.19{\gray 6}\tP 7\\
    \TP{-6}&%
    -3.836{\gray6} \tP{6} &%
     -3.52{\gray 9} \tP 6 &%
    2.5{\gray 6}
    \tP 6&%
    8.84{\gray 3}  \tP 6&%
    -3.06{\gray 6}\tP 6\\
    \TP{-5}&%
    -3.782{\gray 9} \tP{5}&%
     -3.42{\gray 7} \tP 5&%
    2.6{\gray 5}
    \tP 5&%
    8.91{\gray 4} \tP 5&%
    -2.89{\gray 5} \tP 5\\ %
    \TP{-4}&%
    -3.707{\gray 7} \tP{4}&%
     -3.28{\gray 9} \tP 4&%
     2.7{\gray
      5}\tP 4&%
    9.00{\gray 5}\tP 4&%
    -2.66{\gray 1} \tP 4\\
    \TP{-3}&%
    -3.598{\gray 3} \tP{3}& %
    -3.08{\gray 8} \tP 3 &%
    2.9{\gray
      2}\tP 3&%
    9.14{\gray 6}\tP 3&%
    -2.32{\gray 2}\tP 3\\
    \TP{-2}&%
    -3.420{\gray 6} \tP{2}&%
     -2.762{\gray 9} \tP 2&%
    3.1{\gray
      3}\tP 2&%
    9.36{\gray 5} \tP 2&%
    -1.77{\gray 7} \tP 2\\
    \TP{-1}&%
    -3.071{\gray 5} \tP{1}&%
     -2.139{\gray 3} \tP 1&%
    3.6{\gray
      2}\tP 1&%
    9.78{\gray 8}\tP 1&%
    -7.43{\gray 9}\\
   1\phantom{0^{-0}}&%
    -1.978{\gray 3} \tP{0}&%
     -2.99{\gray 3} \tP
    {-1}&%
    4.9{\gray 3}&%
    11.03{\gray 1}&%
    2.18{\gray 0}\\
    \TP 1\phantom{{}^{-}}&%
    -1.068{\gray 3} \tP{-4}&%
     .999{\gray
      5}&%
    1.18{\gray 1}&%
    1.60{\gray 6}&%
    2.06{\gray 6}\\
    \TP 2\phantom{{}^{-}}&-2.{\gray 8} \tP{-44}& %
    .99999{\gray
      9}&%
    1.00{\gray 2} & %
     1.00{\gray 8}&%
    2.00{\gray 1}\\
\hline 
  \end{array}\) 
  \caption{Numerical energies\protect\footnotemark for several \(U(1)\) sector states at
    \(N=3\)} 
\end{table}

\footnotetext{In the table, the last digit (grayed out) is indicative
  and is not claimed to be accurate.}}

As seen in the figure \ref{fig:N=3_spec}, this method allows to  
compute numerically the energies of excited states of the \(U(1)\) sector  for  the whole physically interesting range of lengths \(L\), from deep IR to deep UV region.  
We can see that in the IR, \(\L\to \infty\) limit, the energies of individual  states basically tend to the number of ``particles'' forming the state - the number of the Bethe roots \(\th_j\): The vacuum energy tends to  
\(0\), while the energies of the states \(\th_0,\th_1\) and \(\th_2\) tend to  
1, and the energy of \(\th_{00}\) tends to \(2\). In the conformal  
limit \(\L\to 0\), we will see that the behavior is defined by the
``particle's'' mode numbers:   
The energy goes to \(\frac{2\pi}{\L}(-\frac  {N^2-1}{12}+n)\) where \(N^2-1\) is the conformal central charge (the number of free bosonic fields of the model in this regime) and   
\(n\) is the total momentum mode number (see the discussion of IR and  UV limits in the next section).

\afterpage{\begin{figure}        
  \centering        
\includegraphics{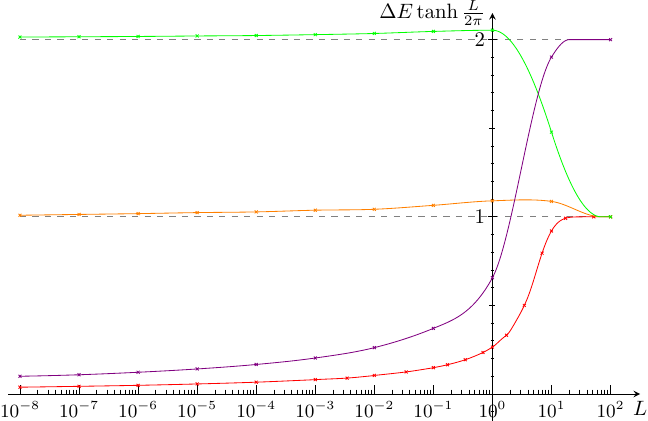}        
  \caption{Differences of energies, as functions of \(\L\),  at \(N=3\). For low lying excited states, the combination \((E-E_{vacuum})\tanh{\frac L{2\pi}}\) is plotted, as in figure \ref{fig:N=3_spec}.}
  \label{fig:N=3_dif}
\end{figure}}

\paragraph{Numerical restrictions}  

As the length \(\L\) is decreasing, the algorithm looks worse and  
worse converging, and the densities become more and more peaked around the endpoints of the distribution. By  
choosing a small enough interpolation step (the   
densities \(f_j\) are numerically defined by polynomial interpolation from a  
finite number of values), it was nevertheless possible to make the  
algorithm reasonably convergent for the considered states and lengths \(\L\), when \(N=3\). Decreasing further the interpolation step  means  
increasing the computation time and the  necessary amount of memory, which puts a practical limit to our precision and to the minimal length%
    .  

Unfortunately, at \(N\geq 4 \) the calculations become heavier and 
with the size of interpolation steps we can afford  our algorithm becomes 
instable already   for \(\L\) of order \(\sim 1 \) (which means we  
cannot really check,  for instance, the conformal limit). At the moment we  
cannot say whether this instability has a physical meaning (like some symmetry  
breakdown, or some new type of singularity appearing) or whether  
it is just a numerical artifact, due to a poor numerical accuracy, or to  
the choice of the equations. For instance, it could be that the  
equation we iterate  stops to correspond to a contraction mapping but still 
has a fixed point, and maybe even that, by rewriting slightly the
functions, it could become a contraction again, and extracting its
fixed point would be possible by iterations.  %

\section{IR and UV limits}

In this section, we will compute analytically the IR, finite size corrections  to particular lowest lying states, as well as the UV, small size limit for a general state of the model.  These results are very useful for checking our numerical data.  
    
\subsection{Leading order results at large \texorpdfstring{$\L$}{L}}        
\label{sec:Luscher}

The approach of this paper allows to compute the first exponential finite size correction, the so called L\"uscher correction, to the  
energy at large \(\L\), as we will  show now on a few examples.  

\subsubsection{Vacuum}        
    
The large \(\L\) behavior of vacuum is given by the condition that         
\begin{eqnarray}        
\left.        
  Y_{a,0}\right|_{\substack{\L\to\infty\\L\ddot{u}scher}}&=&(T_{a,1})^2        
  e^{-L \tp_a}        
\end{eqnarray}        
where \(T_{a,1}\) is equal, according to the formulas \eqref{eq:DetDef}, \eqref{eq:U1SpChF} and \eqref{eq:U1SpChL}, to the binomial coefficient \(        
\left(        
  \begin{array}{c}        
    N\\a        
  \end{array}        
\right)\). This is obtained from \eqref{eq:largeLY1} where
\(\varphi=1\),  and can be plugged directly into \eqref{eq:vacMenergy} to 
get the energy to the leading order. For instance, if \(N=3\), one gets
\(E_{vacuum}\simeq -9\sqrt{\frac 2{\pi L}} e^{-L}\). By construction, this expression fits well our numerical results when  
\(L\) is large enough\footnote{
When  \(\L\geq 4\), the energy
deviates from the asymptotic behaviour 
precision by less than \(10\%\), and this deviation quickly decreases
when \(L\) increases.%
}.  
    
\subsubsection{Mass Gap at \texorpdfstring{$N=3$}{N=3}}        
When \(N=3\) it suffices to compute \(Y_{1,0}\) to get the energy,        
because \(Y_{2,0}=\overline{Y_{1,0}}\).        
    
Moreover the previous analysis shows that        
\begin{eqnarray}        
  Y_{1,0}&=& e^{-\L \cosh(\frac{2 \pi} 3 \th )        
      }  \frac{(T_{1,1})^2}        
      {T_{0,0}T_{2,0}} \frac 1 {{S_0(\th-3\iq)}^2}\frac        
  {\varphi(\th-3\iq)}{\varphi(\th+\iq)}\frac 1 {\chi_{_{CDD}}(\th-3\iq)}\\        
&=& e^{-\L \cosh(\frac{2 \pi} 3 \th )        
      }  \frac{(3 \th -5\iq)^2}        
      {\left(\th+\iq\right)^2} \frac 1 {{S_0(\th-3\iq)}^2}\frac 1 {\chi_{_{CDD}}(\th-3\iq)}        
\end{eqnarray}        
that enables to compute at large \(\L\) the leading order value of the        
integral term in (\ref{eq:oddNU1energy}).         
    
Unlike  the vacuum case, we have to  compute  now the second term of (\ref{eq:oddNU1energy})
which is a bit tricky  as it  
involves the   
position of the Bethe root. This position can be estimated by  
computing the densities to the leading order, to deduce the first  
correction to \(T_{1,0}\) in order to solve the equation \(T_{1,0}(\th_0  
+i/4)=0\).  

For the mass gap, this root should be at the origin, up to exponential
corrections in \(L\). Moreover, one can show\footnote{These large \(L\) 
  expressions are obtained by neglecting integral terms in the  
  determinant expression of \(T_{1,0}\).} that \(T_{1,0}(0+i/4)\sim \frac i
6 f_2(0)+i f_3(0)=\mathcal{O}(e^{-L \sqrt{3}/2})\), while
\(T_{1,0}'(0+i/4)\sim i\), so that \(T_{1,0}(\th_0  
+i/4)=0\) gives \(\th_0\sim - {\frac 1
  6}f_2(0)-f_3(0)\). Using the asymptotic  
expression for \(f_j\)'s (which can be extracted by keeping only the
leading order in \(T_{a,s}\) and in \(d_{a,j}\) in the formula 
\eqref{eq:closed}), one gets  
\(\theta_0   
 \sim \frac{i e^{-\sqrt{3} L/2}        
  \Gamma\left(-\frac{1}{3}\right)^2        
  \Gamma\left(\frac{2}{3}\right)^2}{\sqrt{3} \pi        
  \Gamma\left(\frac{1}{3}\right)^2}  
\), so that the second term in (\ref{eq:oddNU1energy}),        
which is \(\sinh \left(\frac{2 \pi }{3} \left(\theta_0        
      -i/4\right)\right)-\sinh \left(\frac{2\pi}{3}         
    \left(\bar\theta_0        
      +i/4\right)\right)\nonumber         
  \) can be computed at leading        
order.        
    
That gives 
\begin{equation} 
\label{eq:MGLuscher} 
  E^{mass gap}_{\L\to\infty}\simeq 1 - \left(\frac{32 e^{-\sqrt 3 L/2} \pi^3}{ \Gamma\left( \frac  
  1 3\right)^6}\right) 
\end{equation} 
which is in very good agreement the  
numerical results, as can be seen in fig. \ref{fig:MGgraph}. 

Moreover, this expression \eqref{eq:MGLuscher} coincides exactly with
the so-called \(\mu\)-term~\cite{Luscher:1985dn,Klassen:1990ub}, which is known to dominate the finite-size
corrections in the presence of  bound states.

 \subsection{Conformal limit at \texorpdfstring{$\L\to 0$}{L->0}}   

Let us start from the vacuum. At very small \(\L\), the effective   
coupling constant becomes very small  \(e^{2}_{0}(\L)\simeq\frac{2\pi   
}{|\log \L|} \) and we can linearize  the field on the group manifold   
in the vicinity of \(g(\s,\tau)= I\)   as \( g^{-1}\partial_\mu g
\simeq %
{i}\,\partial_\mu A \), where \(A(\sigma,\tau)\) is a Hermitian
\(N\times N\) traceless matrix field.   The \(SU(N)\) PCF model should
become a 2d CFT of \(N^2-1\)  massless bosons: \(R(\L)\)  is very big,
the action  \eqref{eq:ActionDef} becomes
\begin{equation}\label{O4sigma}   
\mathcal{S} =%
\frac{1}{2e_{0}^{2}(\L)} \int {\mathrm{d}\tau} \,\int_0^L{\mathrm{d}\sigma} \:\ \sum_{\a=1}^{2}{\rm tr} %
{\left[\vphantom{h_{}^{_{}}}\right.} (\partial_{\alpha} A)^{2}%
{\left.\vphantom{h_{}^{_{}}}\right]}\quad +\,O\left(e^{4}_{0}(\L)\right).   
 \end{equation}   

In the ground state, the Casimir effect will dominate the limiting energy: \(E_{0}\simeq-\frac{\pi c}{6\L}+O(1/\log ^{4}(\L^{-1}))\), with the central charge \(c=N^2-1\), which gives  \(E_{0}\frac{\L}{2\pi}\simeq-\frac{N^2-1}{12}\). {}   

The energies of excited states are   
\begin{equation}\label{CFTlev}   
\frac{\L}{2\pi}E_{\vec n_{1}\vec n_{2}\vec n_{3}\cdots }(\L)\simeq -\frac{
  (N^2-1) }{12}   
+\sum _{k=1}^{\cN}\sum_{\alpha=1}^{N^2-1}| n_{k}^{(\alpha)}|   
\end{equation}   
where   
\(\vec n_{k}=(n_k^{(1)},n_k^{(2)},\cdots,n_k^{(N^2-1)})\) are the momentum numbers of components of the \(k\)'th particle  and \(\cN\) is
the number of particles constituting the state.   
We see that the small \(\L\) asymptotics of our plots are well  
described by this formula. The vacuum, and the states \(\th_0\),  
\(\th_{0,0}\), \(\ldots\) have total momentum zero, and their energy satisfies  
\(\frac{\L}{2\pi}E(\L)\simeq -\frac{
  (N^2-1) }{12} \).
This formula explains  
well the fact that the corresponding plots on fig.\ref{fig:N=3_spec}  
converge, though slowly, as inverse logarithm of \(\L\), to \(-(N^2-1)/12 \).   
 On the other  
hand, a state like \(\th_1\) has the momentum number equal to \(1\) and  
\(\frac{\L}{2\pi}E(\L)\simeq -\frac{
  (N^2-1) }{12}+1\), etc.  

The approximate   behavior of the  states  \(\theta_{0},\theta_{00}\), etc, at very small \(\L\)'s can be explained by the fact that the quantum fields are dominated by their zero modes. Since the momentum modes are not excited the field \(g(\sigma,\tau)\) does not depend on \(\sigma\).   
The action and the hamiltonian become: %
\begin{equation}   
\mathcal{S%
}\approx %
{%
  \frac 1 2} \frac{%
  \L}{e_{0}^{2}(\L)} \int d\tau\, \,{\rm tr}(g^{-1}\partial_{\tau} g)^{2},\qquad \hat H= \, \frac{e_{0}^{2}(\L)}{2\L}\,{\rm tr}\hat  J^2   
 \end{equation}   
  where  the  \(g(\tau) \) represents the coordinate of a material  
  point (a top) on the group manifold, and \(\hat J\) is the  
  corresponding angular momentum operator.  The quantum mechanical  
  spectrum of this system is well known:    
  the quantum states are classified according to the irreducible  
  representations of \(su(N)\) characterized by  highest weight with  
  components \((m_1\geq m_2\geq,\cdots,\geq m_N) \) usually represented  
  by a Young tableaux \(\lambda\) with \(N\) rows with the lengths  
  \(m_j,\,\,j=1,\cdots,N\). The operator \(\,{\rm tr}\hat  J^2    
 \) is nothing but the second Casimir operator with the well known  
 eigenvalues, so that      \begin{equation}\label{ZEROMODE}    
\frac{\L}{2\pi}(E_{\lambda}-E_{0})\approx\frac{1}{4\pi}e_{0}^{2}(\L) \,{\rm tr}\hat  J^2    
  \,=\frac{e_{0}^{2}(\L)}{4\pi} \sum_{k=1}^Nr_k(r_k-2k+N+1)\;   
\end{equation}   
where \(r_k=m_k-\frac{1}{N}\sum_{j=1}^Nm_j\).

 We can use  the two-loop expression  for our length  scale  \(L\ll1\)\(\)   \begin{eqnarray*}
 \frac{L}{c}  =\sqrt{\frac{4\pi}{N }} \,\,\frac{1}{e_0}\,e^{-\frac{4\pi}{Ne_0^2 }} \end{eqnarray*} where the constant \(c\) is defined by the renormalization scheme (corresponding to our TBA approach). We will use this constant as a fitting parameter in our numerical results. This gives for the two-loop running coupling: \(\frac{4\pi}{N e_0^2}=\log\frac{c}{L}+\frac{1}{2}\log\log\frac{c}{L}\). 

  For instance, for a  
state with only \(M\) real roots in the asymptotic limit (and without  
self-conjugated complexes of roots), we have \(m_1=M\), \(m_{k\geq 2}=0\),  
and hence   
\begin{equation}\label{SymZEROMODE}    
\frac{\L}{2\pi}(E_{\theta_{{\underbrace{\{0,0,\dots,0\}}_{M\,\text{times}}  
    }}}-E_{0})\approx\frac{e_{0}^{2}(\L)}{4\pi N} (N-1)M(M+N)  \,.
\end{equation}

\begin{figure}        
  \centering        
\includegraphics{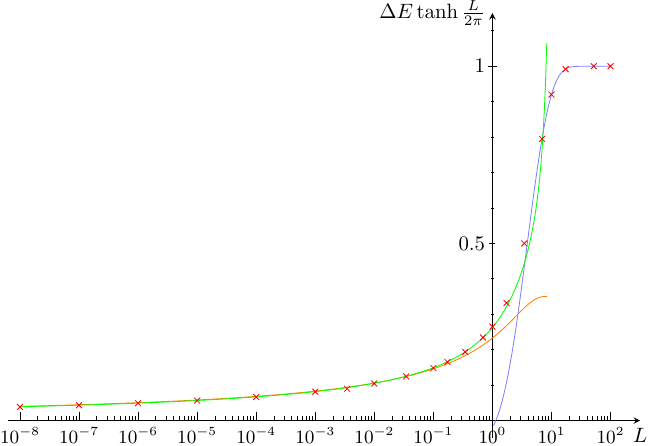}
  \caption{Mass gap \(\Delta E=E_{\theta_0}-E_{vacuum}\). The numeric results (red crosses) are compared to the
 analytic L\"uscher correction \eqref{eq:MGLuscher}  for \(E^{mass gap}_{\L\to\infty}\) [blue curve], to the 1-loop expression 
\(\frac{\L}{2\pi}[E_{\theta_{0}}(\L)-E_{0}(\L)]\approx\frac{8 }{9}\frac{1}{\log \frac{c}{L}}\) [orange curve], and to the 2-loop expression 
\(\frac{\L}{2\pi}[E_{\theta_{0}}(\L)-E_{0}(\L)]\approx\frac{8 }{9}\frac{1}{\log \frac{c}{L}+\frac{1}{2}\log\log \frac{c}{L}}\) [green curve] \eqref{smallLmassgap},  where
 \(c\) is  
chosen as the best fit for the \(L<10^{-1}\) data\(^{\ref{ft:c}}\)%
.}
  \label{fig:MGgraph}        
\end{figure}

 The 2-loop perturbative calculation of the  
mass gap \([E_{\theta_{0}}(\L)-E_{0}(\L)]\) for  \(\L\ll1\) was done in  
\cite{Shin:1996gi}.  It  was compared with the numerical results  
following from the TBA approach in \cite{Balog:2003yr} for \(N=2\).  Here we cite this result only  
  for the mass gap (\(M=1)\), in the logarithmic approximation  using the 2-loop result of  \cite{Mana:1996nz,Mana:1996pk}:     

\begin{equation}\label{smallLmassgap}   
\frac{\L}{2\pi}[E_{\theta_{0}}(\L)-E_{0}(\L)]\approx\frac{N^2-1 }{N^2}\frac{1}{\log \frac{c}{L}+\frac{1}{2}\log\log \frac{c}{L}},\qquad(\L\ll 1)   
\end{equation}which is in the perfect agreement with \eqref{ZEROMODE},  
as well as with our numerics, as seen from the Fig~\ref{fig:MGgraph}. In this figure the red and green curve show respectively the one and two-loop expressions of the mass gap when \(N=3\). The value of \(c\) used in this picture is chosen to fit the \(L<10^{-1}\) numeric data\footnote{\label{ft:c}Explicitly we used the value \(c=44\) for the one-loop best-fit, and \(c=17\) for the two-loop best-fit.}%
, and remarkably enough, the two-loop expression is reasonably close to the exact result up to \(L\simeq 3\).

  In principle, the three loop running coupling is also   known  in a certain scheme \cite{Rossi:1993zc,Rossi:1994yp} but accounting for it will be beyond the accuracy of our numerics.

The formula \eqref{ZEROMODE} also gives a prediction for the state 
\(\th_{00}\) with two zero-momentum-particles, namely, for \(N=3\) we have 
\(\frac{E_{\theta_{00}}(\L)-E_{0}(\L)}{E_{\theta_{0}}(\L)-E_{0}(\L)}\simeq 
\frac {5} 2\). This result matches our numerics when
\(\L\) is smaller than \(1\) (see the fig. \ref{fig:N=3_dif}), up to the
precision announced in table \ref{tab:new}.

Although the motivation of our approach was based on adding some terms, such
as  resolvents \(F_k\) in \eqref{eq:qFinitSizeU1F}-\eqref{eq:qFinitSizeU1L}, correcting 
the infinite size solution, it reproduces correctly these conformal 
expressions, which shows that this description is not only  
accurate in some vicinity of \(\L=\infty\), but even in the conformal  
limit where \(\L\) is very small. It proves  that these terms were  
added by us into the ansatz  \eqref{eq:qFinitSizeU1F}-\eqref{eq:qFinitSizeU1L} in a sufficiently general manner to describe the relevant exact  
solutions of the \(Y\)-system at any finite \(L\).  

\section{Discussion}        

We have presented here, on the example of the \(SU(N)\) principal
chiral field model, a   powerful and rather general
approach to  the study of  finite volume   spectrum of various
integrable 1+1 dimensional sigma-models. The approach continues the
ideas of \cite{Gromov:2008gj} where the method was proposed on the
example of the \(SU(2)\) PCF, but for \(N>2\)   the method has to be
seriously reconsidered due to many new physical features
w.r.t. the \(N=2\) case. In particular, the presence of the bound state
particles and the non-reality of the Bethe roots at finite \(L\) show
a few qualitatively new features within our approach.  

  For virtually all integrable sigma models at a finite volume, the TBA-like  approach initiated
by   Al.Zamolodchikov can be summarized in a very universal system of
  functional equations, the \(Y\)-system.  The Y-system is equivalent  
to the famous Hirota equation - the Master equation of integrability 
describing in this case the integrable discrete dynamics with respect 
to a pair of ``representational'' variables, \(a,s\) and the spectral 
parameter (rapidity) \(\theta\). The boundary conditions for \(a,s\)
are defined  
by the symmetry algebra of the model, whether as the analytic 
structure w.r.t. the \(\theta\) variable is in general the most 
complicated issue, largely defining the dynamics of the model. However,  
 in fact even the possible analyticity structures are greatly 
constrained by Hirota dynamics and by the symmetry algebra. It would 
be interesting to   classify possible types  of analyticity stemming 
from Hirota dynamics and some simple physical arguments (relativistic 
invariance, crossing, absence of certain singularities, etc.) related
to the finite volume sigma models, similarly to the S-matrix bootstrap
theory of Al.\(\&\)A.Zamolodchikov valid only at infinite volume.   
This could lead to an interesting classification of sigma models
themselves and  possibly to the discovery of new integrable  models.  It would also help  bypassing the standard TBA approach, poorly justified and,  strictly speaking, valid only for the vacuum state.    
  
In this paper, we managed to transform the finite volume spectral problem for one such  
relativistic  \(\sigma\)-model, the \(SU(N)\times SU(N)\) principal
chiral field into a finite set of NLIEs. It was achieved by solving the underlying finite \(\L\)
Y-system in terms of   
Wronskian determinants of a finite number of Q-functions  and
parameterizing these Q-functions by \(N-1\) densities correcting their
large \(\L\) asymptotics to any  finite \(\L\).

Our work generalizes the analytic and numerical results of  
\cite{Gromov:2008gj} to \(N\geq 2\), and we could numerically check, at  
least when \(N=2, 3\), that this procedures solves the \(Y\)-system, and  
enables to compute energies for a wide range of lengths \(\L\), 
compatible with the UV conformal limit and the IR finite size 
(L\"uscher) corrections.  On the way, we conjectured a natural 
generalization of the energy formula for excited states  for the 
\(U(1)\) at finite \(\L\), to \(N\geq 2\). This generalization appears 
to be unexpectedly non-trivial and looks different for even and odd 
\(N\). The question of  definition of the energy formula for excited 
states deserves a better understanding and hopefully the eventual 
derivation.   

The analysis was done for \(U(1)\) sector states, and it  
certainly can and should  be generalized to any excited state, as was done in   
\cite{Gromov:2008gj} for \(N= 2\).  
\label{general_excited}    
 To do this, one will  
have to understand the asymptotic terms and the structure of  
zeroes. In particular, some extra zeroes should appear in Y-functions which might  
affect the way the energy is computed by contour manipulation. Apart  
from that, the main difference with \(U(1)\) sector should be that for  
non-symmetric states (i.e. when \(Y_{a,s}\neq Y_{a,-s}\)), it will be  
necessary to introduce \(N-1\) densities for the right wing and \(N-1\)  
densities for the left wing. One would have to write  
\eqref{eq:YcentralR} as \(2(N-1)\) different equations, by writing the  
left-hand side either as  
\(\frac{T_{1,1}^{(R)}T_{1,-1}^{(R)}}{T_{2,0}^{(R)}T_{0,0}^{(R)}}\) or as  
\(\frac{T_{1,1}^{(L)}T_{1,-1}^{(L)}}{T_{2,0}^{(L)}T_{0,0}^{(L)}}\). Our approach based on the Wronskian solution of Y-system should
a priori still enable us to compute the energies of   
these states. 

An interesting problem which our approach might help to solve is the
planar \(N\to\infty\) limit in PCF at finite \(L\). This PCF model has a rich history of
its comparison to QCD and it might provide an important example of exactly solvable 2+1 dimensional bosonic string theory, similarly to
the matrix quantum mechanical model of  the 1+1 dimensional, \(c=1\)
non-critical string theory proposed and solved in
\cite{Kazakov:1988ch}. The exact and explicit solution for this limit was given
in the case of infinite volume \(L\) but in the presence of a specific
"magnetic" fields \cite{Fateev:1994ai}. The finite volume solution might provide a deeper understanding of 't~Hooft limit in asymptotically free QFT's and even reveal some new physical phenomena, such as a possible large \(N\) phase transition at some \(L_c\), in analogy with the   Yang-Mills theory on the 2D sphere \cite{Douglas:1993iia} (equivalent to the one-dimensional PCF).  This, seemingly 2nd order, phase transition was already observed numerically in \cite{Campostrini:1994ih}.

As concerns the numerics, our algorithm  
converges very well for any length when \(N\leq 3\), but  for \(N\geq  
4\) it is very unstable for small enough  length \(\L\), already at
\(\L\lesssim1\) (which means for instance that we   
cannot really check the convergence to  the conformal UV
limit). Hopefully this instability has no direct physical meaning and
is just a numerical artifact, due to a poor numeric accuracy or to   
the bad choice of the iteration procedure for our NLIEs.  It would be good to compare our results with the high precision Monte-Carlo simulations of \(SU(3)\) \cite{Mana:1996nz,Mana:1996pk} for the mass gap as a function of the volume, but these papers are mostly concerned with reaching  the infinite volume asymptotically free regime for the latice PCF model with the  torus, rather than cylindric boundary conditions.

We believe that this method of derivation of a finite system of NLIEs for
integrable sigma models is  general and powerful enough to work for
much more complicated cases of  AdS/CFT correspondence, such as the
superstring on the   AdS\(_5\times S^5\) background dual to N=4 SYM
theory, and the so called ABJM model where the \(Y \)-system was
already discovered
\cite{Gromov:2009tv,Bombardelli:2009xz,Gromov:2009at}. The Wronskian
quasiclassical character \cite{Gromov:2010vb} and even the full
quantum solution of the Hirota  dynamics for AdS\(_5\)/CFT\(_4\)
\cite{GKLT} are already available. The understanding of the
very rich and complicated analyticity structure of Q-functions for
short operators is of a great help for the derivation of the AdS/CFT NLIE.

\section*{Acknowledgments}

The work  of VK was partly supported by  the ANR grants INT-AdS/CFT
(BLAN-06-0124-03)  and   GranMA (BLAN-08-1-313695) and the grant RFFI
08-02-00287. 
We would like to thank B.Vicedo for the participation on an early
stage of this work,   M.~Douglas, A.~Hegedus,  E.~Sobko, A.~Tsvelik, P.~Wiegmann,
A.~Zabrodin for useful discussions,  L.~Hollo  for carefully reading this version of the  manuscript, and especially  N.Gromov and
P.Vieira for their constant attention and numerous important comments
in the course of this work. VK also thanks Simons Center for geometry
and physics (Stony Brook) and Nordita institute (Stockholm) where a
part of this work was done, for their kind hospitality.   

\section*{Notes for version 2}

Compared to the version~1, the present version fixes a few typos,
and notably improves the precision of numerical results. A part of this
improvement comes from a better treatment of the \emph{mode numbers}
of the Bethe roots, described in a new paragraph in sect.\ref{sec:FSBE}. We also added at the end of sect.\ref{sec:q-ansatz} the
twisted version of our Ansatz, allowing to generalize these results to
the twisted boundary conditions.

\end{document}